\documentclass[11pt,a4paper]{article}
\usepackage{graphicx}
\usepackage{bm}
\usepackage{slashed}
\usepackage{braket}
\usepackage[dvipsnames]{xcolor}
\usepackage{amsmath, dsfont}
\usepackage{amsfonts}
\usepackage{amssymb}
\usepackage{setspace}
\usepackage[colorlinks=true
  ,urlcolor=blue
  ,anchorcolor=blue
  ,citecolor=blue
  ,filecolor=blue
  ,linkcolor=blue
  ,menucolor=blue
  ,pagecolor=blue
  ,linktocpage=true
  ,pdfproducer=medialab
  ,pdfa=true
]{hyperref}

\usepackage{hypcap}
\usepackage{soul}
\usepackage{jcappub}
\setstcolor{blue}
\newcommand{\lgr}{\mathcal{L}}

\newcommand{\Mpl}{M_{\rm Pl}}
\newcommand{\tMpl}{\tilde{M}_{\rm Pl}}
\newcommand{\D}{\textnormal{d}}

\newcommand{\tA}{\tilde{A}({\tilde{\chi}})}
\newcommand{\tAv}{\tilde{A}(v_{\tilde{\chi}})}
\newcommand{\tAvp}[1]{\tilde{A}{#1}(v_{\tilde{\chi}})}
\newcommand{\tAvexp}[1]{\tilde{A}^{#1}(v_{\tilde{\chi}})}

\newcommand{\edit}[1]{{#1}}
\title{A particle's perspective on screening mechanisms}
\author[a]{Sergio Sevillano Mu\~noz}
\affiliation[a]{Institute for Particle Physics Phenomenology, Department of Physics, Durham University, \\Durham DH1 3LE, U.K.
}
\emailAdd{sergio.sevillano-munoz@durham.ac.uk}
\date{}
\preprint{IPPP/24/45}
\abstract{Screening mechanisms are a natural method for suppressing long-range forces in scalar-tensor theories as they link the local background density to their strength. Focusing on Brans-Dicke theories, those including a non-minimal coupling between a scalar degree of freedom and the Ricci scalar, we study the origin of these screening mechanisms from a field theory perspective, considering the influence of the Standard Model on the mechanisms. Additionally, we further consider the role of scale symmetries on screening, demonstrating that only certain sectors, those obtaining their mass via the Higgs mechanism, contribute to screening the fifth forces. This may have significant implications for baryons, which obtain most of their mass from the gluon's binding energy. \edit{However, a definitive statement requires extending these calculations to bound states. We show that the non-minimally coupled field's interactions with the Higgs lead to an extensive region of the parameter space where screening mechanisms create spatially dependent fermion masses}. We say that the field over-screens when this effect is more significant than the fifth forces suppressed by screening mechanisms, as we illustrate for the chameleon and symmetron models.}
\begin{document}

\maketitle
\newpage
\section{Introduction}\label{sec:introduction}
Current cosmological and particle physics tensions motivate the extension of the models we use to describe our universe. While the usual route is to modify the Standard Model by adding Beyond the Standard Model terms, a popular alternative is to extend the gravitational sector accordingly. For this, there are different choices one can take, but only some of them are consistent with both General Relativity and Quantum Field Theory formalism and symmetries. Among these, we will focus on theories that allow direct couplings between an additional scalar degree of freedom and curvature terms, those called scalar-tensor theories~\cite{Yasunori:Fujii_2003}.

These types of couplings are said to be \textit{non-minimal} and, without symmetries to prevent them, they will always be generated by the RG flow of any theory defined at a given specific scale~\cite{Herranen:2014cua, Markkanen:2018bfx,Steinwachs:2011zs}; this connects these theories to UV completing theories such as string theory~\cite{Cicoli:2023opf}. The number of possible such non-minimal couplings can be reduced by considering only those that generate up to second-order differential equations of motion, collected into the Horndeski theory~\cite{Horndeski:1974wa,Kobayashi:2019hrl} (which, in turn, can be generalised into Beyond-Horndeski~\cite{Traykova:2019oyx,Gleyzes:2013ooa} or DHOST theories~\cite{Langlois:2015cwa,Langlois:2018dxi}). The main example for such theories is the Brans-Dicke theory~\cite{Brans:1961sx}, where the dynamical scalar field couples to the Ricci scalar in the gravitational action, controlling the strength of gravity through its vacuum expectation value (vev). Popular examples of such models include those in which the Higgs is non-minimally coupled to gravity, as is required in Higgs inflation or the Higgs-dilaton theory~\cite{Wetterich:1987fm, Buchmuller:1988cj, Shaposhnikov:2008xb, Shaposhnikov:2008xi, Blas:2011ac, Garcia-Bellido:2011kqb, Garcia-Bellido:2012npk, Bezrukov:2012hx, Henz:2013oxa, Rubio:2014wta, Karananas:2016grc, Ferreira:2016vsc, Ferreira:2016kxi, Casas:2017wjh, Ferreira:2018qss}. However, performing such extensions to gravity will lead to new interactions that must be considered. In particular, the dominant interactions will normally be long-ranged, through the so-called \textit{fifth forces}.  Different experiments and tests have been performed to constrain the strength of these fifth-forces, varying from the stability of compact objects~\cite{Damour:1992we, doneva2022scalarization,Cardoso:2013opa}, the evolution of the universe~\cite{Avilez:2013dxa}, dynamics inside the Solar System~\cite{Bertotti:2003rm,Fischer:2024eic} or Equivalence Principle tests~\cite{EotWash,Merkowitz:2010kka} (see Ref.~\cite{Burrage:2017qrf} for a review on all tests). Altogether, these studies point out that our universe is accurately described by the standard Einstein-Hilbert action. This requires fine-tuning to suppress any existing fifth forces so that they are ${\sim 10^{-5}}$ times smaller than gravity~\cite{Bertotti:2003rm, Williams:2005rv}.

Among all possible scalar-tensor theories, some of them present screening mechanisms that relate the fifth-force strength to the local density of the background. For example, this can be obtained by increasing the effective mass of the mediating field, as for the chameleon~\cite{Burrage:2017qrf, Khoury:2003rn,Burrage_2016}, or by decreasing its coupling constant to matter, as for the symmetron~\cite{Hinterbichler:2010es,Hinterbichler:2011ca}. This would be a natural explanation for the lack of evidence of these theories at large scales. 
It is, therefore, no surprise that one of the main current motivations in this field is to find phenomenological implications that evade or \textit{see through} screening mechanisms.  One popular way of doing so is studying these theories from a particle physics perspective. For example, from a Quantum Field Theory approach, one can test for the consistency of the interactions by imposing unitarity and causality to be maintained~\cite{deRham:2021fpu}, or study the effect of these new interactions within atoms~\cite{Brax:2022olf} and quantum systems~\cite{Burrage:2018pyg,Kading:2023mdk}. Similarly, given the neutrality of the non-minimally coupled field, it has been proposed that the production of such external states in scattering processes would lead to effective missing energy signals that can be measured in colliders~\cite{Brax_2009,Argyropoulos:2023pmy,SevillanoMunoz:2023loj}. To carry out all these calculations, it is common to use the fact that extensions of gravity can always be expressed as effective Beyond the Standard Model physics~\cite{Burrage:2018dvt,Copeland:2021qby}. In Ref.~\cite{Burrage:2023eol}, the authors explore the emergence of screening mechanisms from this Beyond the Standard Model description, making a series expansion of the coupling functions for simplicity. Here, we extend their work by not assuming this series expansion and studying the impact of different mass-generating mechanisms on the screening of fifth forces. Throughout this work, we find that screening mechanisms create a space-dependent \edit{fermion masses}. This would effectively present itself as a mismatch of the gravitational and inertial masses, which is very well constrained locally, allowing for a deviation of $\Delta(M_G/M_I)=(-2.0\pm2.0)\times 10^{-13}$~\cite{Williams_2004}.

This paper is structured as follows: In section~\ref{sec:BD as BSM}, we show how scalar-tensor theories can be expressed as Beyond the Standard Model theories, where we also calculate the emergence of screening mechanisms in a toy model Lagrangian of QED+Abelian-Higgs. In section~\ref{sec: Mass gen}, we generalise the calculations to allow for multiple fermionic states and different scale-breaking mechanisms. Then, in section~\ref{sec: over-screening}, we study the impact of popular screening mechanisms such as the chameleon~\cite{Burrage:2017qrf, Khoury:2003rn,Burrage_2016} and symmetron models~\cite{Hinterbichler:2010es,Hinterbichler:2011ca} inside \edit{a toy model for} galactic disks. Finally, we conclude in section~\ref{sec: conclusion}.

\section{Screening mechanisms in the Standard Model}\label{sec:BD as BSM}
From all possible extensions of the Einstein-Hilbert action, we focus on the Brans-Dicke theory~\cite{Brans:1961sx}, where a non-minimally coupled scalar field ($\varphi$) allows for a varying effective gravitational constant, which will be defined through its vacuum expectation value. This theory takes the following generic form:
\begin{equation}\label{eq:BDgeneric}
	S=\int \D^4 x \sqrt{-g}\left[-\frac{F(\varphi)}{2}R +\frac{1}{2} g^{\mu\nu}\partial_\mu \varphi \partial_\nu \varphi - U(\varphi) + \lgr_{\rm m}\{\psi_i,g_{\mu\nu}\}\right],
\end{equation}
where $R$ is the Ricci scalar, built with the Jordan frame metric $g^{\mu\nu}$, $\lgr_{\rm m}\{\psi_i\}$ is the matter Lagrangian containing the fields $\psi_i$, and $F(\varphi)$ and $U(\varphi)$ are functions controlling the non-minimal coupling to gravity and the potential for the scalar field, respectively. These types of actions are more generic than they seem, as gravitational sectors containing functions of the Ricci scalar terms ($f(R)$ theories) can be effectively expressed as Brans-Dicke theories~\cite{DeFelice:2010aj}. A priori, the only requirement that these theories must satisfy is that, at late times, the field $\varphi$ is constant or very slowly evolving through the potential, such that a constant effective Planck mass can be generated via
\begin{equation}
	\Mpl^2\equiv F(\varphi)|_{\varphi\to v_\varphi},
\end{equation}
where $v_\varphi$ is the vev of the field. Current bounds estimate that the allowed rate of change for the Planck mass is constrained to $\dot{\Mpl}/\Mpl\leq(-2\pm7)\cdot10^{-13}$/year~\cite{Muller:2007zzb}, where $\dot{\Mpl}=\D \Mpl/\D t$.

Depending on the chosen action, we expect new channels of interactions to emerge in the matter sector indirectly through the modification of gravity. These new channels can be expressed as Beyond the Standard Model interactions either by staying in the Jordan frame and linearising and canonically normalising gravity, as in Refs.~\cite{Copeland:2021qby,SevillanoMunoz:2022tfb,SevillanoMunoz:2023loj}, or by taking a conformal transformation to the Einstein frame, where the Ricci scalar absorbs the non-minimal coupling~\cite{Burrage:2018dvt}.  Here, we will take the second route, which is the standard way to study modified theories of gravity. In terms of the generic action in Eq.~\eqref{eq:BDgeneric}, this conformal (or Weyl) transformation is given by
	\begin{align}\label{eq:Weyl transformations}
		g_{\mu\nu}\to &\frac{\tMpl^2}{F(\varphi)}\tilde{g}_{\mu\nu}, &
		g^{\mu\nu}\to &\frac{F(\varphi)}{\tMpl^2}\tilde{g}^{\mu\nu},
	\end{align}
	where $\tilde{g}_{\mu\nu}$ and $\tMpl$ are the Einstein frame metric and constant Planck mass, respectively. Substituting this transformation into Eq.~\eqref{eq:BDgeneric}, the gravitational action now takes the following form:
	\begin{align}
		S=\int \D^4x\sqrt{-\tilde{g}} \bigg[-&\frac{\tMpl^2}{2}\tilde{R} + \frac{\tMpl^2}{2}\left[\frac{1}{F(\varphi)}+\frac{3F'(\varphi)^2}{2F^2(\varphi)}\right]\tilde{g}^{\mu\nu} \partial_\mu \varphi \partial_\nu \varphi \nonumber \\
		-&\frac{\tMpl^4}{F^2(\varphi)} U(\varphi) +  \tilde{\lgr}_{\rm m}\{\psi_i,\tilde{g}_{\mu\nu},\varphi\}\bigg],\label{eq:BD EinsteinFrame}
	\end{align}
	where we can see that while we recover a canonical gravitational action, given by the Einstein frame Ricci scalar $\tilde{R}$, all the modifications appear in the matter Lagrangian, $\lgr\{\psi_i,\tilde{g}_{\mu\nu},\varphi\}$, through the minimal couplings of the fields to gravity. To work with this action, first, we need to redefine the scalar field $\varphi$ so that it is canonically normalized, solving
	\begin{equation}\label{eq:canon to chi}
	 \chi=\int^\varphi_{\varphi_0} \D \hat\varphi \,\tMpl\sqrt{\frac{1}{F(\hat\varphi)} +\frac{3F'(\hat\varphi)^2}{2F^2(\hat\varphi)}},
	\end{equation}	
	where $\varphi_0$ is a constant that can be taken to be zero without loss of generality. Substituting this into Eq.~\ref{eq:BD EinsteinFrame}, we obtain
	\begin{align}
		S=\int \D^4x\sqrt{-\tilde{g}} \bigg[-\frac{\tMpl^2}{2}\tilde{R} + \frac12\tilde{g}^{\mu\nu} \partial_\mu  \chi \partial_\nu  \chi -V( \chi) +  \tilde{\lgr}_{\rm m}\{\psi_i,\tilde{g}_{\mu\nu}, \chi\}\bigg],\label{eq:BD EinsteinFramecanon}
	\end{align}
	 where $V( \chi)\equiv \tMpl^4 U(\varphi)/F^2(\varphi) $. Now, we need to define a matter sector, for which we will choose a toy model of Abelian-Higgs+QED, following the work in~\cite{SevillanoMunoz:2023loj, SevillanoMunoz:2022tfb}, as it can be easily extended to more complicated Standard Model sectors. Thus, we will work with the following Jordan frame matter action,
	 	\begin{align}\label{ActionJordanFull}
	 	S_{\rm m}[g_{\mu\nu}]=\int \D^4{x} \sqrt{-g} &\left[-\frac{1}{4}g^{\alpha\mu}g^{\beta\nu}F_{\alpha\beta}F_{\mu\nu}+\frac{1}{2}g^{\mu\nu}\partial_\mu \phi \partial_\nu \phi \right. \nonumber \\ 
	 	&+i\bar{\psi}e^\mu_a\gamma^{a}\partial_\mu\psi +\frac{1}{2}\bar{\psi}e^\mu_a\gamma^{a}\Omega_\mu\psi-q\bar{\psi}e^\mu_a\gamma^{a}A_\mu\psi \nonumber \\
	 	&-y\bar{\psi}\phi\psi +\frac{1}{2}\mu^2 \phi^2 -\frac{\lambda}{4!}\phi^4 -\frac{3\mu^4}{2\lambda}\bigg],
	 \end{align}
	 where $\phi$ is a scalar field corresponding to the would-be Higgs field, $\psi$ is a fermionic field and $A_\mu$ the massless U(1) gauge field, with its strength tensor $F_{\mu\nu}=\partial_\mu A_\nu - \partial_\nu A_\mu$.  Additionally, we have explicitly expressed the minimal couplings to gravity; particularly, we included the vierbeins ${e^a}_\mu$ and the spin connection
	 $\Omega_\mu$,
	 that ensures that the kinetic energy of the fermion is kept Hermitian and symmetric under conformal transformations. 
	 
	 Applying the Weyl transformation to the Einstein frame from Eq.~\eqref{eq:Weyl transformations} to this matter action and appending the gravitational sector from Eq.~\eqref{eq:BD EinsteinFramecanon} leads to 
	 \begin{align}\label{ActionEinsteinFull}
	 	S[\tilde{g}_{\mu\nu}]=\int \D^4{x} \sqrt{-\tilde{g}} \nonumber&\left[-\frac{\tMpl^2}{2}\tilde{R} + \frac12\tilde{g}^{\mu\nu} \partial_\mu  \chi \partial_\nu  \chi -V( \chi)\right.\nonumber\\
	 	& -\frac{1}{4}\tilde{g}^{\alpha\mu}\tilde{g}^{\beta\nu}F_{\alpha\beta} F_{\mu\nu}+\frac{A^2(\chi)}{2}\tilde{g}^{\mu\nu}\partial_\mu \phi \partial_\nu \phi +A^3(\chi)i\bar{\psi}\tilde{e}^\mu_a\gamma^{a}\partial_\mu\psi \nonumber \\ 
	 	&+ A^3(\chi)\frac{1}{2}\bar{\psi}\tilde{e}^\mu_a\gamma^{a}\psi\left(\tilde\Omega_\mu +3i \frac{A'(\chi)}{A(\chi)}\partial_\mu \chi\right)-A^3(\chi)q\bar{\psi}\tilde{e}^\mu_a\gamma^{a}A_\mu\psi  \nonumber \\
	 	&-A^4(\chi)y\bar{\psi}\phi\psi +A^4(\chi)\left(\frac{1}{2}\mu^2 \phi^2 -\frac{\lambda}{4!}\phi^4 -\frac{3\mu^4}{2\lambda}\right)\bigg],
	 \end{align}
	where $A^2(\chi)=\tMpl^2/F(\varphi(\chi))$ and all tilded quantities are built with the Einstein frame metric. To canonically normalise the matter fields, we can redefine each field depending on its scaling dimension by
	\begin{align}
		\phi\to&A^{-1}(\chi)\tilde{\phi}, & \psi\to&A^{-3/2}(\chi) \tilde{\psi},
	\end{align}
	from which we obtain
	 \begin{align}
		S[\tilde{g}_{\mu\nu}]=\int \D^4{x} \sqrt{-\tilde{g}} &\left[-\frac{\tMpl^2}{2}\tilde{R} + \frac12\tilde{g}^{\mu\nu} \partial_\mu  \chi \partial_\nu  \chi -V( \chi)\right.\nonumber\\
		& -\frac{1}{4}\tilde{g}^{\alpha\mu}\tilde{g}^{\beta\nu}F_{\alpha\beta} F_{\mu\nu}+\frac{1}{2}\tilde{g}^{\mu\nu}\partial_\mu\tilde{\phi} \partial_\nu \tilde{\phi} \nonumber \\ 
		& -\frac{A'(\chi)}{A(\chi)}\tilde{\phi}\tilde{g}^{\mu\nu}\partial_\mu\chi \partial_\nu \tilde{\phi}+\frac{\tilde{\phi}^2}{2}\frac{A'(\chi)^2}{A^2(\chi)}\tilde{g}^{\mu\nu}\partial_\mu\chi \partial_\nu \chi \nonumber \\ 
		&+i\bar{\tilde{\psi}}\tilde{e}^\mu_a\gamma^{a}\partial_\mu\tilde{\psi} +\frac{1}{2}\bar{\tilde{\psi}}\tilde{e}^\mu_a\gamma^{a}\tilde\Omega_\mu\tilde{\psi}-q\bar{\tilde{\psi}}\tilde{e}^\mu_a\gamma^{a}A_\mu\tilde{\psi} \nonumber \\
		&-y\bar{\psi}\tilde{\phi}\psi +\left(\frac{1}{2}\mu^2A^{2}(\chi)\tilde{\phi}^2 -\frac{\lambda}{4!}\tilde\phi^4 -A^4(\chi)\frac{3\mu^4}{2\lambda}\right)\bigg].\label{eq: BD EF rescaled}
	\end{align}
	We can see that the fermionic and gauge fields have absorbed the couplings to the non-minimally coupled field. However, while the gauge sector is completely free of fifth forces, that will not be the case for the fermionic field, as it will still become affected through the Yukawa interaction to the would-be Higgs, which has now obtained kinetic and mass mixing couplings to the new scalar (third and last line of Eq.~\eqref{eq: BD EF rescaled}, respectively). Given that the only field directly coupling to the fifth forces is the would-be Higgs, these theories are equivalent to Higgs-portal theories~\cite{Burrage:2018dvt}. Some of these couplings between $\tilde{\phi}$ and $\chi$ can be absorbed by canonically normalising $\chi$ into
    \begin{equation}\label{eq:canon chi to tildechi}
		\tilde{\chi}=\int \D\hat{\chi} \sqrt{1+\tilde{\phi}^2\left(\frac{A'(\hat{\chi})}{A(\hat{\chi})}\right)^2},
	\end{equation}
	which does not lead to important modifications as it is usually the case ${\tilde{\phi}A'(\chi)}/{A(\chi)}\ll1$. With this, the Einstein frame action takes the following form:
	\begin{align}\label{ActionEinsteinCanonical}
		S[\tilde{g}_{\mu\nu}]=\int \D^4{x} \sqrt{-\tilde{g}} &\left[-\frac{\tMpl^2}{2}\tilde{R} + \frac12\tilde{g}^{\mu\nu} \partial_\mu   \tilde{\chi} \partial_\nu   \tilde{\chi} -\tilde{V}(  \tilde{\chi})\right.\nonumber\\
		& -\frac{1}{4}\tilde{g}^{\alpha\mu}\tilde{g}^{\beta\nu}F_{\alpha\beta} F_{\mu\nu}+\frac{1}{2}\tilde{g}^{\mu\nu}\partial_\mu\tilde{\phi} \partial_\nu \tilde{\phi} \nonumber  -\frac{ \tilde{A}'( \tilde{\chi})}{ \tilde{A}( \tilde{\chi})}\tilde{\phi}\tilde{g}^{\mu\nu}\partial_\mu \tilde{\chi} \partial_\nu \tilde{\phi}\nonumber \\ 
		&+i\bar{\tilde{\psi}}\tilde{e}^\mu_a\gamma^{a}\partial_\mu\tilde{\psi} +\frac{1}{2}\bar{\tilde{\psi}}\tilde{e}^\mu_a\gamma^{a}\tilde\Omega_\mu\tilde{\psi}-q\bar{\tilde{\psi}}\tilde{e}^\mu_a\gamma^{a} {A}_\mu\tilde{\psi} \nonumber \\
		&-y\bar{\psi}\tilde{\phi}\psi +\left(\frac{1}{2}\mu^2 \tilde{A}^{2}( \tilde{\chi})\tilde{\phi}^2 -\frac{\lambda}{4!}\tilde\phi^4 - \tilde{A}^4( \tilde{\chi})\frac{3\mu^4}{2\lambda}\right)\bigg],
	\end{align}
	where $\tA\equiv A(\chi)$ and $\tilde{V}(\tilde{\chi})\equiv V(\chi)$. The kinetic mixing channel is subdominant in the modifications of the matter Lagrangian~\cite{Burrage:2018dvt}, so we will focus on the couplings in the potential of $\tilde{\phi}$. 
 
    Now that the fields are canonically normalised, we can perturb the scalar fields around their vacuum expectation values, which will introduce the screening effects into the theory. This calculation has been done in the limit where $\tAv\approx1$ in Ref.~\cite{Burrage:2023rqx}, which constrains the allowed range of values for $v_{\tilde{\chi}}$ depending on the chosen potential and coupling function $\tA$. Here, we will keep $\tAv$ generic, studying the full phenomenological implications of screening mechanisms on the matter Lagrangian. To find the scalar fields' vacuum expectation values, we need to solve the following system of equations
	\begin{equation}\label{eq:vevsEDO}
	\begin{split}
		 \frac{\lambda}{6}v_{\tilde{\phi}}^4-\mu^2v_{\tilde{\phi}}^2&\tilde{A}^2(v_{\tilde\chi}) +\rho_\psi=0,\\
		  \tilde{A}'(v_{\tilde{\chi}})\tilde{A}^3(v_{\tilde{\chi}})\frac{6\mu^4}{\lambda}-\mu^2v_{\tilde{\phi}}^2&\tilde{A}'(v_{\tilde{\chi}})\tilde{A}(v_{\tilde{\chi}})+\tilde{V}'(v_{\tilde\chi})=0,
	\end{split}
\end{equation}
where $\tAvp{'}=\D\tA/\D \tilde{\chi}|_{\tilde{\chi}\to v_{\tilde{\chi}}}$, $V'(v_{\tilde{\chi}})=\D V({\tilde{\chi}})/\D \tilde{\chi}|_{\tilde{\chi}\to v_{\tilde{\chi}}}$, and $\rho_\psi=yv_{\tilde{\phi}}\bar{\psi}\psi$ is the local energy density of the fermion fields and will be the term introducing the screening mechanisms. For generic $\tAv$ and $\tilde{V}(\tilde{\chi})$, this gives
\begin{align}
	&v_{\tilde{\phi}}^2=\frac{v^2}{2}\tilde{A}^2(v_{\tilde{\chi}})\left(1+\sqrt{1-\frac{4\rho_\psi}{v^2\mu^2\tAvexp{4}}}\right) \nonumber \\ &\tilde{V}'(v_{\tilde{\chi}})=-\frac{\tAvp{'}}{\tAv}\rho_\psi,\label{eq:vevs}
\end{align}
where $v^2=6\mu^2/\lambda$. Although for all practical scenarios we can take $\rho_\psi\ll v^2\mu^2\tAvexp{4}$,\footnote{\edit{Note that the local variation of $\tAv$ with the background density may cause this limit to be invalid. In such a case, the stable minimum of the Higgs field would be at $v_{\tilde{\phi}} = 0$, reunifying the electroweak sector into one force. This would significantly affect Standard Model phenomenology as all fermions would become massless. We will ignore that limit in $\tAv$ in this work.}} keeping the square root terms in $v_{\tilde{\phi}}$ is important to obtain the right result for $v_{\tilde{\chi}}$, as the leading order terms cancel in the derivation (as pointed out in Ref.~\cite{Burrage:2023rqx}). This result agrees with Ref.~\cite{Burrage:2023rqx} when making a series expansion of $\tA$ and assuming $\tAv\approx1$. However, contrary to their conclusions, we find that allowing for an arbitrary value for $\tAv$ clearly shows that the particle description leads to the same effective potential for $\tilde{\chi}$ as in the standard calculations for screening mechanisms (i.e., $\tilde{V}_{\rm eff}’(\tilde{\chi})= \tilde{V}’(\tilde{\chi}) + \tilde{A}’(\tilde{\chi})\rho_\psi/\tilde{A}(\tilde{\chi})$). This means that the Higgs field does not interfere with screening mechanisms but just enables them.

To see the effects of screening mechanisms on the matter Lagrangian, we have to shift the fields around their vevs ($\tilde{\phi}\to \tilde{\phi}+v_{\tilde{\phi}}$ and $\tilde{\chi}\to \tilde{\chi}+v_{\tilde{\chi}}$). Although we do not take a series expansion of $\tAv$ to calculate the vacuum expectation values (i.e., we don't use $\tAv\approx1$), we will assume that the field's perturbations around this vev are suppressed, which is the case for the most popular screening models. This allows us to make the following series expansion of $\tA$ around the vacuum expectation value,
\begin{equation}\label{eq: A(chi+vev)}
	\tilde{A}(\tilde{\chi}+v_{\tilde{\chi}})=\tilde{A}(v_{\tilde{\chi}})+\tilde{A}'(v_{\tilde{\chi}}) \chi +\tilde{A}''(v_{\tilde{\chi}})\frac{\chi^2}{2}+\dots .
\end{equation}
Performing this expansion to Eq.~\eqref{ActionEinsteinCanonical}, we obtain
	\begin{align}\label{ActionEinsteinAfterVev}
	S[\tilde{g}_{\mu\nu}]=\int \D^4{x} \sqrt{-\tilde{g}} &\left[-\frac{\tMpl^2}{2}\tilde{R} + \frac12\tilde{g}^{\mu\nu} \partial_\mu   \tilde{\chi} \partial_\nu   \tilde{\chi} -\tilde{V}(  \tilde{\chi}+v_{\tilde{\chi}})\right.\nonumber\\
	& -\frac{1}{4}\tilde{g}^{\alpha\mu}\tilde{g}^{\beta\nu}F_{\alpha\beta} F_{\mu\nu}+\frac{1}{2}\tilde{g}^{\mu\nu}\partial_\mu\tilde{\phi} \partial_\nu \tilde{\phi} \nonumber  -\frac{ \tilde{A}'( \tilde{\chi})}{ \tilde{A}( \tilde{\chi})}\tilde{\phi}\tilde{g}^{\mu\nu}\partial_\mu \tilde{\chi} \partial_\nu \tilde{\phi}\nonumber \\ 
	&+i\bar{\tilde{\psi}}\tilde{e}^\mu_a\gamma^{a} \tilde{\nabla}_\mu\tilde{\psi}  -y\bar{\psi}\tilde{\phi}\psi-yv_{\tilde{\phi}}\bar{\psi}\psi +H(\tilde{\phi},\tilde{\chi})\bigg],
	\end{align}
	where $H(\tilde{\phi},\tilde{\chi})$ is the vev-shifted Higgs potential, which is important for calculating the coupling strength of the long-range mode to fermions and its associated mass. However, before calculating these quantities, let us focus on an important consequence of the environmentally dependent vev for $\tilde{\chi}$. This has to do with the generated mass of the fermions, which from Eq.~\eqref{ActionEinsteinAfterVev} is given by 
	\begin{equation}\label{eq: mass fermion}
		m_\psi=yv_{\tilde{\phi}}\approx y\sqrt{\frac{6\mu^2}{\lambda}}\tAv.
	\end{equation}
	In the $\tAv\approx1$ case, this would evaluate to the constant $m_\psi\approx y\sqrt{6\mu^2/\lambda}$. However, allowing for any value for $\tAv$ leads to big implications for screening mechanisms, \edit{as depending on the potential and coupling function $\tA$, the masses of fermionic fields can be very sensitive to changes in the background's energy density}. Since measurements are always made in comparison to other measurements (equivalent to using a ruler for distances), not all physics will be affected by this mass shift.  Nonetheless, forces charged by particles' masses (i.e., gravity and fifth forces) will experience this mass shift, as they can be made dimensionless observables by comparing them to the electric force, which does not pick up any mass correction. \edit{Similarly, it is possible to create dimensionless observables by taking the ratio of experiments carried out in different backgrounds, which would show the mass shift of elementary particles. We demonstrate in Appendix~\ref{Appendix A} how this same result would be obtained working directly in the Jordan frame.} Given that $\tAv$ is background-dependent, this implies that screening mechanisms will seem as if the strength of gravity varies with space, following the background distribution of matter. This can be used to constrain scalar-tensor theories against both the cosmological evolution of $\tilde{A}(\tilde{\chi})$~\cite{Wang:2012kj} and the local evolution of the Planck mass~\cite{Muller:2007zzb}.
 
	For completeness, we will also obtain the effective $\tilde{\chi}$ field mass and fifth-force coupling due to screening using a generic $\tAv$ and $\tilde{V}(\tilde{\chi})$. As earlier, we will assume that the perturbations are suppressed, which is justified for most models, and we will use the following substitutions for this demonstration:
	\begin{align}
		a=&\tAv, & b=&\tAvp{'}, & c=&\tAvp{''},
	\end{align}
	such that Eq.~\eqref{eq: A(chi+vev)} can be expressed as
	\begin{equation}
		\tilde{A}(\tilde{\chi}+v_{\tilde{\chi}})=a+b\chi +\frac{c}{2}\chi^2+\cdots .
	\end{equation}
	With this, when expanding the Higgs potential around the vevs, we obtain
	\begin{align}
		H(\tilde{\phi},\tilde{\chi})=&\frac{\mu^2}{2} \left(a^2 \tilde{\phi}^2 +2ab v_{\tilde{\phi}}\tilde{\phi}\tilde{\chi}+(b^2+ac)v_{\tilde{\phi}}^2\chi^2\right)\nonumber \\
		-&\frac{\lambda}{4}v_{\tilde{\phi}}^2\tilde{\phi}^2-\frac{3\mu^4}{2\lambda}(2a^3c+6a^2b^2)\tilde{\chi}^2+\dots,
	\end{align}
	where the ellipsis includes cubic and higher-order terms. Including the potential $\tilde{V}(\tilde{\chi}+v_{\tilde{\chi}})$, we find that the mass matrix for this Lagrangian is given by
 \begin{equation}
     M=\begin{pmatrix}
         m_{\tilde{\phi}}^2 &  \quad
        &\alpha_{\tilde{\phi\chi}}\\\\
         \alpha_{\tilde{\phi\chi}} &  \quad & m_{\tilde{\chi}}^2
        \end{pmatrix},
 \end{equation}
    with
	\begin{align}
		m_\phi^2=&-a^2\mu^2+\frac{\lambda}{2}v_{\tilde{\phi}}^2\nonumber\\
		m_\chi^2=&-\mu^2(b^2+ac)v_{\tilde{\phi}}^2\chi^2+\frac{v^2\mu^2}{2}(2a^3c+6a^2b^2)-\tilde{V}''(v_{\tilde{\chi}}),\nonumber\\
            \alpha_{\tilde{\phi\chi}}=&\mu^2 ab v_{\tilde{\phi}}.
	\end{align}
	We cannot make the usual addition of terms in the would-be Higgs mass into $m_\phi^2=2a^2\mu^2$, as its vev now has subleading $\rho_\psi$ dependence that will be important when calculating the fifth-force strength, see Eq.~\eqref{eq:vevs}. To isolate the long-ranged from the Higgs interaction, we need to canonically diagonalise the mass matrix, as in Refs.~\cite{Burrage:2018dvt, Burrage:2023rqx}. After some algebra, we find two modes of propagation of forces, which have the following masses
	\begin{align}
		m^2_h=&2\mu^2a^2,\nonumber\\
		m^2_\sigma=&\tilde{V}''(v_{\tilde{\chi}}) + \frac{c}{a}\rho_\psi.\label{eq: meff}
	\end{align}
	Here, we find one heavy mode, corresponding to the Higgs boson, $h$, and one light mode transmitting the long-range forces, $\sigma$. As we can see, the Higgs boson also has an environmentally dependent mass proportional to $a=\tAv$. Moreover, we find that the screened mass of the light mode also agrees with the standard case, where it is given by $m_\sigma^2=V_{\rm eff}''(v_{\tilde{\chi}})$, with $V_{\rm eff}(\tilde{\chi})=V(\tilde{\chi})+\tilde{A}^{'}(\tilde{\chi})\rho_\psi/\tA$.
	
	This equivalence with the standard screening description extends to the coupling of fifth forces to the matter fields. For this, we need to find how the fields $\tilde{\phi}$ and  $\tilde{\chi}$ transform into the canonical fields $h$ and $\sigma$. Since the fifth forces are introduced through the fermion's Yukawa coupling to the would-be Higgs, we only need to study how the following term in the matter Lagrangian transforms
	\begin{equation}
		\lgr\supset -y\bar{\psi}\tilde{\phi}\psi.
	\end{equation}
	After some algebra, we find that the light mode channel couples to the fermions as follows
	\begin{equation}
		\lgr\supset -yv \tAvp{'}\bar{\psi}\sigma\psi,
	\end{equation}
	where, using $\rho_\psi=yv_{\tilde{\phi}}\bar{\psi}\psi=yv\tAv\bar{\psi}\psi$, we can express the fifth force coupling to energy density as
	\begin{equation}
		\lgr\supset -\frac{\tAvp{'}}{\tAv}\sigma\rho_\psi,
	\end{equation}
	agreeing with the standard result. 
	
	\section{Screening for different mass generation mechanisms}\label{sec: Mass gen}
	So far, we have seen the effect of one single fermionic field on the screening of the non-minimally coupled field and on the mass of elementary particles. Here, we will extend these results into three different contexts: First, we will consider multiple fermion fields with masses sourced either from the Higgs mechanism or from explicit mass terms; then, we will explore the case for classical scale invariant theories whose scale symmetry breaks dynamically. Additionally, we will study the implications of screening for baryons, which are known to obtain most of their mass from the gluons' binding energy. 
	\subsection{Multiple sources of screening}
	As a generic example for this case, we will consider the following Jordan frame matter action
	\begin{align}
	S_{\rm m}[g_{\mu\nu}]=\int\D x^4\,\sqrt{-g}\bigg[&\frac{1}{2}g^{\mu\nu}\partial_\mu\varphi\partial_\nu\varphi+\frac{1}{2}g^{\mu\nu}\partial_\mu\phi\partial_\nu\phi \nonumber\\
	+&\left(\sum_\alpha^N i\bar{\psi_\alpha}\gamma^{\mu}{\nabla}_\mu\psi_\alpha \nonumber -y_\alpha\bar{\psi_\alpha}\phi\psi_\alpha\right)\\
	+&\left(\sum_\beta^M i\bar{\tau_\beta}\gamma^{\mu}{\nabla}_\mu\tau_\beta \nonumber -m_{\beta,0}\bar{\tau_\beta}\tau_\beta\right)\\
	-&U(\varphi)+\frac{1}{2}\mu^2\phi^2-\frac{\lambda}{4!}\phi^4-\frac{3\mu^4}{2\lambda}\bigg],
\end{align}
which contains the would-be Higgs providing mass to $N$ number of fermions $\psi_\alpha$, a family of $M$ fermions $\tau_\beta$, which have an explicit mass term, $m_{\beta,0}$, and the non-minimally coupled field $\varphi$ with its potential $V(\varphi)$. For completeness, one would have to also add the U(1) gauge field, but that sector trivially decouples from $\varphi$, as we saw in the last section.

Taking the conformal transformation to the Einstein frame as given in Eq.~\eqref{eq:Weyl transformations}, we obtain that the matter action transforms as
\begin{align}
	S_{\rm m}[\tilde{g}_{\mu\nu}]=\int\D x^4\,&\sqrt{-\tilde{g}}\bigg[\frac{1}{2}\tilde{g}^{\mu\nu}\partial_\mu\chi\partial_\nu\chi+A^2(\chi)\frac{1}{2}\tilde{g}^{\mu\nu}\partial_\mu\phi\partial_\nu\phi\nonumber\\
	+&\left(\sum_\alpha^N iA^3(\chi)\bar{\psi}_\alpha\tilde{e}^\mu_a\gamma^{a} \big(\tilde{\nabla}_\mu+\frac32\partial_\mu\log(A(\chi))\big)\psi_\alpha \nonumber -y_\alpha A^4(\chi)\bar{\psi}_\alpha\phi\psi_\alpha\right)\\
	+&\left(\sum_\beta^M iA^3(\chi)\bar{\tau}_\beta\tilde{e}^\mu_a\gamma^{a} \big(\tilde{\nabla}_\mu+3\partial_\mu\log(A(\chi))\big)\tau_\beta \nonumber -m_{\beta,0 }A^4(\chi)\bar{\tau}_\beta\tau_\beta\right)\\
	-&V(\chi)+A^4(\chi)\left(\frac{1}{2}\mu^2\phi^2-\frac{\lambda}{4!}\phi^4-\frac{3\mu^4}{2\lambda}\right)\bigg],
\end{align}
where we remind that the all tilded quantities are built using the Einstein frame metric, $\tilde{g}_{\mu\nu}$. As in section~\ref{sec:BD as BSM}, to canonically normalise all fields' kinetic energies, we must redefine them depending on their scaling dimension as
\begin{align}
	\phi\to&A^{-1}(\chi)\tilde{\phi}, & \psi_i&\to A^{-3/2}(\chi) \tilde{\psi}_i, & \tau_i&\to A^{-3/2}(\chi) \tilde{\tau}_i,
\end{align}
leading to
\begin{align}
	S_{\rm m}[\tilde{g}_{\mu\nu}]=\int\D x^4\,\sqrt{-\tilde{g}}\bigg[&\frac{1}{2}\tilde{g}^{\mu\nu}\partial_\mu\tilde{\chi}\partial_\nu\tilde{\chi}+\frac{1}{2}\tilde{g}^{\mu\nu}\partial_\mu\tilde{\phi}\partial_\nu\tilde{\phi}-\frac{ \tilde{A}'( \tilde{\chi})}{ \tilde{A}( \tilde{\chi})}\tilde{\phi}\tilde{g}^{\mu\nu}\partial_\mu \tilde{\chi} \partial_\nu \tilde{\phi}\nonumber\\
	+&\left(\sum_\alpha^N i\bar{\tilde{\psi}}_\alpha\tilde{e}^\mu_a\gamma^{a} \tilde{\nabla}_\mu\tilde{\psi}_\alpha \nonumber -y_\alpha \bar{\tilde{\psi}}_\alpha\phi\tilde{\psi}_\alpha\right)\\
	+&\left(\sum_\beta^M i\bar{\tilde{\tau}}_\beta\tilde{e}^\mu_a\gamma^{a} \tilde{\nabla}_\mu\tilde{\tau}_\beta \nonumber -m_{\beta,0} \tilde{A}(\tilde{\chi})\bar{\tilde{\tau}}_\beta\tilde{\tau}_\beta\right)\\
	-&\tilde{V}(\tilde{\chi})+\left(\frac{1}{2}\mu^2\tA^2\tilde{\phi}^2-\frac{\lambda}{4!}\tilde{\phi}^4-\frac{3\mu^4}{2\lambda}\tA^4\right)\bigg],
\end{align}
where we have already canonically normalised the $\chi$ field into $\tilde{\chi}$ as in Eq.~\eqref{eq:canon chi to tildechi}. Here, we see that those fermionic fields that acquire their mass via the Higgs mechanism will contribute to the screening of the fifth forces through their interactions with the would-be Higgs (as in section~\ref{sec:BD as BSM}), while the ones containing an explicit mass will interact directly with the non-minimally coupled field. The question is: will both groups of fermions contribute equally to the screening of $\chi$ and pick up the same background-dependent mass?

The answer does not require a very complicated computation, as we only need to calculate the vacuum expectation values for the scalar fields as for section~\ref{sec:BD as BSM}. Defining the local densities of each category of fermions as $\rho_\psi=\sum_\alpha^N y_\alpha v_{\tilde{\phi}}\bar{\tilde{\psi}}_\alpha\tilde{\psi}_\alpha$ and $\rho_\tau=\sum_\beta^M m_{\beta,0} \bar{\tilde{\tau}}_\beta \tilde{\tau}_\beta$, we obtain
	\begin{equation}\label{eq:vevsEDO case2}
	\begin{split}
		\frac{\lambda}{6}v_{\tilde{\phi}}^4-\mu^2v_{\tilde{\phi}}^2&A^2(v_{\tilde\chi}) +\rho_\psi=0,\\
		\tilde{A}'(v_{\tilde{\chi}})\tilde{A}^3(v_{\tilde{\chi}})\frac{6\mu^4}{\lambda}-\mu^2v_{\tilde{\phi}}^2&\tilde{A}'(v_{\tilde{\chi}})\tilde{A}(v_{\tilde{\chi}})+\tilde{V}'(v_{\tilde\chi})+\frac{\tAvp{'}}{\tAv}\rho_\tau=0.
	\end{split}
\end{equation}
Comparing these equations with Eq.~\eqref{eq:vevsEDO}, we can see that the system of equations is equivalent under the replacement $\tilde{V}'(v_{\tilde{\chi}})\to\tilde{V}'(v_{\tilde{\chi}})+\rho_\tau \tAvp{'}/{\tAv}$. Making these same substitutions into the results from the one field case in Eq.~\eqref{eq:vevs}, we find
\begin{align}\label{eq:vevs case 2}
	v_{\tilde{\phi}}^2=&\frac{v^2}{2}\tilde{A}^2(v_{\tilde{\chi}})\left(1+\sqrt{1-\frac{4\rho_\psi}{v^2\mu^2\tAvexp{4}}}\right)  \nonumber\\ \tilde{V}'(v_{\tilde{\chi}})=&-\frac{\tAvp{'}}{\tAv}(\rho_\psi+\rho_\tau),
\end{align}
where we have again used $v^2=6\mu^2/\lambda$. We can see that although the would-be Higgs density corrections depend only on the energy density of the fields it interacts with ($\psi_\alpha$), both groups of fermions contribute equally to the screening of the non-minimally coupled field. Moreover, expanding the function $\tA$ around the field's vev as
\begin{equation}
	\tilde{A}(\tilde{\chi}+v_{\tilde{\chi}})=\tAv+\tAvp{'}\tilde{\chi}+\tAvp{''}\frac{\tilde{\chi}^2}{2}+\dots,
\end{equation}
we find that the masses of the fermion fields are given by
\begin{align}
	m_\alpha=&y_\alpha v_{\tilde{\phi}}=y_\alpha v \tAv\nonumber\\
	m_\beta=&m_{\beta,0} \tAv,
\end{align}
meaning that once a field contributes to the screening of the non-minimally coupled field, it will always experience the same shift in its mass.
\subsection{Scale symmetry breaking mechanisms and screening effects}
Scale symmetries and the mechanisms breaking them have important implications on the strength of fifth forces in scalar-tensor theories. In Refs.~\cite{Garcia-Bellido:2011kqb,Ferreira:2016kxi}, it was shown that a classically scale invariant matter Lagrangian would not couple with the non-minimally coupled field. Additionally, in Refs.~\cite{Burrage:2018dvt,Copeland:2021qby}, this was explored from a field theoretic perspective, working with a matter Lagrangian where the scale symmetry was broken both explicitly (via the Higgs mechanism) and dynamically, by the generation of the vacuum expectation values of the fields, allowing the authors to study the midpoint between both scale generating mechanisms.

In this subsection, we will use the same model as in Refs.~\cite{Burrage:2018dvt,Copeland:2021qby} to see the impact of scale invariance on the screening of the non-minimally coupled field and mass shift of the fermion fields. With this, we will repeat the calculation from section~\ref{sec:BD as BSM} but introduce another scalar singlet, $\theta$, that will help dynamically break the scale symmetry. The scalar sector we will use for this calculation is
\begin{align}
	S_{\rm{SB}}[g_{\mu \nu}]=\int \D^4{x} \sqrt{-g}& \left[\frac{1}{2}g^{\mu \nu}\partial_\mu \phi \partial_\nu \phi +\frac{1}{2}g^{\mu \nu}\partial_\mu \theta \partial_\nu \theta +\frac{1}{2}g^{\mu\nu}\partial_\mu\varphi\partial_\nu\varphi\right.\nonumber\\
	&\quad+\bar{\psi} e^\mu_a \gamma^a \nabla_\mu\psi- y\bar{\psi}\phi\psi -U(\varphi)\nonumber\\ &\quad \left.-W(\phi, \theta)+\frac{1}{2} \mu_{\theta}^{2} A^{-2}(\varphi)\theta^{2}-\frac{\lambda_{\theta}}{4 !} \theta^{4}-\frac{3}{2} \frac{\mu_{\theta}^{4}}{\lambda_{\theta}} A^{-4}(\varphi)\right],
	\label{sm}
\end{align}
where
\begin{equation}
	W(\phi, \theta)=\frac{\lambda}{4 !}\left(\phi^{2}-\frac{\beta}{\lambda} \theta^{2}\right)^{2}-\frac{1}{2} \mu^{2}\left(\phi^{2}-\frac{\beta}{\lambda} \theta^{2}\right)+\frac{3}{2} \frac{\mu^{4}}{\lambda} .\label{comb_pot}
\end{equation}
The limits mentioned earlier correspond to either taking $\beta\to 0$, the explicit scale symmetry breaking leading to a Higgs-like double-well potential, or $\mu\to0$, where we obtain a Higgs-Dilaton potential~\cite{Wetterich:1987fm, Buchmuller:1988cj, Shaposhnikov:2008xb, Shaposhnikov:2008xi, Blas:2011ac, Garcia-Bellido:2011kqb, Garcia-Bellido:2012npk, Bezrukov:2012hx, Henz:2013oxa, Rubio:2014wta, Karananas:2016grc, Ferreira:2016vsc, Ferreira:2016kxi, Casas:2017wjh, Ferreira:2018qss}. In this model, the additional couplings of $\theta$ to the non-minimally coupled field in the last line of Eq.~\eqref{sm} ensure that fifth forces decouple from those terms. Thus, taking the Weyl transformation as usual, we obtain
\begin{align}
	S_{\rm{SB}}[\tilde{g}_{\mu\nu}]=\int \D^4{x} \sqrt{-\tilde{g}}& \left[\frac{1}{2}A^2(\chi)\tilde{g}^{\mu \nu}\partial_\mu \phi \partial_\nu \phi +\frac{1}{2}A^2(\chi)\tilde{g}^{\mu \nu}\partial_\mu \theta \partial_\nu \theta +\frac{1}{2}g^{\mu\nu}\partial_\mu\chi\partial_\nu\chi\right.\nonumber\\
	&\quad+\bar{\psi}A^3(\chi) \tilde{e}^\mu_a\gamma^{a} \big(\tilde{\nabla}_\mu +\frac32 \partial_\mu\log(A(\chi))\big)\psi- yA^4(\chi)\bar{\psi}\phi\psi -V(\chi)\nonumber\\ &\quad \left.-A^4(\chi)W(\phi, \theta)+\frac{1}{2} \mu_{\theta}^{2} A^{2}(\chi)\theta^{2}-\frac{\lambda_{\theta}}{4 !} A^4(\chi)\theta^{4}-\frac{3}{2} \frac{\mu_{\theta}^{4}}{\lambda_{\theta}} \right],
	\label{smEF}
\end{align}
where we have already canonically normalised the non-minimally coupled field into $\chi$, using Eq.~\eqref{eq:canon to chi}. Following the same process as earlier, we now rescale the fields depending on their scaling dimension as
 \begin{align}
 	\phi\to&A^{-1}(\chi)\tilde{\phi} & \theta\to&A^{-1}(\chi)\tilde{\theta}  & \psi&\to A^{-3/2}(\chi) \tilde{\psi},
 \end{align}
leading to
\begin{align}
	S_{\rm{SB}}[\tilde{g}_{\mu\nu}]=\int \D^4{x} \sqrt{-\tilde{g}}&\nonumber \left[\frac{1}{2}\tilde{g}^{\mu\nu}\partial_\mu\tilde{\chi}\partial_\nu\tilde{\chi}+\frac{1}{2}\tilde{g}^{\mu\nu}\partial_\mu\tilde{\phi}\partial_\nu\tilde{\phi}-\frac{ \tilde{A}'( \tilde{\chi})}{ \tilde{A}( \tilde{\chi})}\tilde{\phi}\tilde{g}^{\mu\nu}\partial_\mu \tilde{\chi} \partial_\nu \tilde{\phi}\right.\\
	&\quad+\frac{1}{2}\tilde{g}^{\mu\nu}\partial_\mu\tilde{\theta}\partial_\nu\tilde{\theta}-\frac{ \tilde{A}'( \tilde{\chi})}{ \tilde{A}( \tilde{\chi})}\tilde{\theta}\tilde{g}^{\mu\nu}\partial_\mu \tilde{\theta} \partial_\nu \tilde{\theta}\nonumber\\
	&\quad+\bar{\tilde{\psi}}\tilde{e}^\mu_a\gamma^{a}  \tilde{\nabla}_\mu\tilde{\psi}- y\bar{\tilde{\psi}}\tilde{\phi}\tilde{\psi} -\tilde{V}(\tilde{\chi})\nonumber\\
	&\quad \left.-\tilde{W}(\tilde{\phi}, \tilde{\theta})+\frac{1}{2} \mu_{\theta}^{2} \tilde{\theta}^{2}-\frac{\lambda_{\theta}}{4 !}\tilde{\theta}^{4}-\frac{3}{2} \frac{\mu_{\theta}^{4}}{\lambda_{\theta}} \right],\label{SM chitilde}
\end{align}
where we have also used 
	\begin{equation}\label{eq:canon chi to tildechi SM}
		\tilde{\chi}=\int \D\hat{\chi} \sqrt{1+(\tilde{\theta}^2+\tilde{\phi}^2)\frac{A'(\hat{\chi})^2}{A^2(\hat{\chi})}}.
	\end{equation}
 This leads to the rescaled potential
	\begin{equation}
		\tilde{W}(\tilde{\phi},\tilde{ \theta})=\frac{\lambda}{4 !}\left(\tilde{\phi}^{2}-\frac{\beta}{\lambda} \tilde{\theta}^{2}\right)^{2}-\frac{1}{2} \mu^{2}\tA^2\left(\tilde{\phi}^{2}-\frac{\beta}{\lambda} \tilde{\theta}^{2}\right)+\frac{3}{2} \frac{\mu^{4}}{\lambda}\tA^4 .\label{comb_pot_tilde}
	\end{equation}
	We can see that the $\theta$ sector in the last line of Eq.~\eqref{SM chitilde} has completely decoupled from the $\tilde{\chi}$ field. These couplings have now emerged in $\tilde{W}(\tilde{\phi},\tilde{ \theta})$, which will introduce the fifth forces and the mass shift into the Standard Model. However, in the pure scale-invariant limit ($\mu\to0$), these couplings to the $\tilde{\chi}$ field vanish, leaving the theory free of fifth forces (there would still be kinetic mixings in the Lagrangian, but they do not introduce long-range interactions into the system). 
	
    As in the last section, we need to find the vacuum expectation values for the scalar fields to include the screening effects into the matter Lagrangian. Thus, finding the minimum of the potential with respect to $\tilde{\phi}$, $\tilde{\theta}$ and $\tilde{\chi}$, we obtain
	\begin{equation}\label{eq:vevsEDO SM}
	\begin{split}
		&\frac{\lambda}{6}v_{\tilde{\phi}}^4-\frac{\beta}{6}v_{\tilde{\theta}}^2v_{\tilde{\phi}}^2-\mu^2v_{\tilde{\phi}}^2A^2(v_{\tilde\chi}) +\rho_\psi=0,\\
		&\frac{\beta^2}{6\lambda}v_{\tilde{\theta}}^4-\frac{\beta}{6}v_{\tilde{\theta}}^2v_{\tilde{\phi}}^2-\frac{\beta\mu^2}{\lambda}v_{\tilde{\theta}}^2A^2(v_{\tilde\chi})-\mu_\theta^2 v_{\tilde{\theta}}^2-\frac{\lambda_\theta}{6}v_{\tilde{\theta}}^4=0\\ 
		&\tilde{A}'(v_{\tilde{\chi}})\tilde{A}^3(v_{\tilde{\chi}})\frac{6\mu^4}{\lambda}-\mu^2\left(v_{\tilde{\phi}}^2-\frac{\beta}{\lambda}v_{\tilde{\theta}}^2\right)\tilde{A}'(v_{\tilde{\chi}})\tilde{A}(v_{\tilde{\chi}})+\tilde{V}'(v_{\tilde\chi})=0,
	\end{split}
	\end{equation}
	which, for generic $\tilde{A}(\tilde{\chi})$ and $\tilde{V}(\tilde{\chi})$ gives
	\begin{align}
	&v_{\tilde{\phi}}^2=\frac{1}{2}\left(\frac{\beta}{\lambda}v_{\tilde{\theta}}^2+v^2\tilde{A}^2(v_{\tilde{\chi}})\right)\left(1+\sqrt{1-\frac{2\lambda\rho_\psi}{3\left(\frac{\beta}{6}v_{\tilde{\theta}}+\mu^2\tAvexp{2}\right)^2}}\right) \nonumber \\
	& v_{\tilde{\theta}}^2=\frac{6\mu_\theta^2}{\lambda_\theta},\nonumber\\
	&\tilde{V}'(v_{\tilde{\chi}})=-\frac{\mu^2\tAvp{'}\tAv}{\frac{\beta}{6}v_{\tilde{\theta}}^2+\mu^2\tAvexp{2}}\rho_\psi.\label{eq:vevs SM}
	\end{align}
	We can see that in the explicitly broken limit ($\beta\to0$), we recover the solution in Eq.~\eqref{eq:vevs}, while in the dynamically broken limit ($\mu\to0$), the vev of the non-minimally coupled field completely decouples from the local density. Additionally, once we expand the perturbations around the vacuum expectation value of the fields, we find that the mass for the fermion is given by
	\begin{equation}
		m=yv_{\tilde{\phi}}\approx y\sqrt{\frac{\beta}{\lambda}v_{\tilde{\theta}}^2+\frac{6\mu^2}{\lambda}\tilde{A}^2(v_{\tilde{\chi}})},
	\end{equation}
	showing that in the $\mu\to 0$ limit, we can preserve a massive fermionic sector while avoiding the mass shift due to possible screening of the non-minimally coupled field. This is a property coming from the fact that this model introduces Higgs-portal terms in the Jordan frame to evade fifth forces, which generate a second mass shift that exactly cancels the one generated by the Higgs (see Appendix~\ref{Appendix A} for a demonstration of this fact working directly in the Jordan frame). Thus, matter sectors that break the scale symmetry dynamically will lead to vanishing fifth forces \edit{while generating a fermionic sector of constant, non-zero masses}. 
\label{subsec SM}
\subsection{Baryon contribution to screening mechanisms}
Although baryons are constituted of quarks, these only contribute to a small percentage of the total baryon mass. This is important for the effect that baryonic matter will have on screening mechanisms, as depending on the scale symmetries of the mass components, they will contribute differently to the screening of fifth forces.

The QCD Lagrangian in the Einstein frame for the gravitational sector in Eq.~\eqref{eq:BDgeneric} is given by
\begin{align}
	S_{\rm{QCD}}[\tilde{g}_{\mu\nu}]=\int \D^4{x} \sqrt{-\tilde{g}}&\nonumber \left[\frac{1}{2}\tilde{g}^{\mu\nu}\partial_\mu\tilde{\chi}\partial_\nu\tilde{\chi}+\frac{1}{2}\tilde{g}^{\mu\nu}\partial_\mu\tilde{\phi}\partial_\nu\tilde{\phi}-\frac{ \tilde{A}'( \tilde{\chi})}{ \tilde{A}( \tilde{\chi})}\tilde{\phi}\tilde{g}^{\mu\nu}\partial_\mu \tilde{\chi} \partial_\nu \tilde{\phi}\right.\\
	&\quad-\frac{1}{4}G_{\mu\nu}^aG^{\mu\nu}_a+\bar{\tilde{q}}_a\tilde{e}^\mu_a\gamma^{a}  \tilde{\nabla}_\mu\tilde{q}^a- y_a\tilde{\phi}\bar{\tilde{q}}_a\tilde{q}^a\nonumber\\
	&\quad \left. -\tilde{V}(\tilde{\chi})+\frac{1}{2}\tA^2 \mu^{2} \tilde{\phi}^{2}-\frac{\lambda}{4 !}\tilde{\phi}^{4}-\tA^4\frac{3}{2} \frac{\mu^{4}}{\lambda} \right],\label{SM QCD}
\end{align}
where we have followed the same steps as in previous calculations, but replaced the fermions with the family of $N$ quarks $q_a$, where $a= 1 \to N$, and the U(1) gauge strength tensor $F_{\mu\nu}$ by the gluon strength tensor $G_{\mu\nu}^a$. Notice that the internal gauge group is unimportant in this derivation, just the minimal couplings to gravity, allowing us to extend our previous results without breaking any internal symmetry~\cite{SevillanoMunoz:2022tfb}. 

We can see that, as expected, the non-minimally coupled field does not couple to the gauge fields or fermions' kinetic energies. Therefore, the only fraction of the baryon's mass contributing to the screening of the fields comes from the bare mass of quarks, which is usually subdominant. To get an estimate for the actual coupling of baryons to the screening of the fields, we will focus on the proton's mass, whose renormalised theory contains the contribution from both quarks and gluons by the four-term decomposition~\cite{Metz:2020vxd}
\begin{align}
    M_q=&\langle\bar{\tilde{q}}_a\tilde{e}^\mu_a\gamma^{a}  \tilde{\nabla}_\mu\tilde{q}^a\rangle_R\nonumber\\
    M_g=&\langle\frac{1}{4}G_{\mu\nu\,a}G^{\mu\nu\,a}\rangle_R\nonumber\\
    M_m=&\langle m_a\bar{\tilde{q}}_a\tilde{q}^a\rangle_R\nonumber\\
    M_T=&\langle0\rangle_R,
\end{align}
where $\langle\rangle_R$ is the expectation value at the proton's rest frame. Herein, $M_q$ and $M_g$ correspond to the potential energy of the quarks and gluons, respectively; $M_m$ is the quark mass contribution, which depends on the Higgs vev, by $m_a=y_a v_{\tilde{\phi}}$; and $M_T$ contains the trace anomaly contributions, which exactly cancel for the $T_{00}$ terms of the energy-momentum tensor, those being considered for this calculation (see Ref.~\cite{Metz:2020vxd} for a detailed discussion on the cancellation of the trace anomaly). 

Having an analytical description of the proton's components is still an open problem in physics. However, we can use different methods to estimate the percentage of the proton mass corresponding to the Higgs mechanism. Competing calculations provide different percentages depending on the assumptions and tools used. Following Ref.~\cite{Metz:2020vxd}, we consider the analysis using chiral perturbation theory from Refs.~\cite{Alarcon:2011zs, Alarcon:2012nr, Hoferichter:2015hva} and Lattice QCD~\cite{PhysRevD.102.054517}, where up to, and including, charm quarks are considered. In the first scenario, they obtain that the quark mass contribution to the proton mass is of $0.074\pm0.08\,{\rm GeV}$, equivalent to ~$8.0\pm8.5\%$, while using Lattice QCD a contribution of $0.187 \pm 0.023\,{\rm GeV}$ is obtained, corresponding to around $20\pm25\%$ of the proton's total mass.\footnote{Had we considered the trace anomaly for this calculation, we would find a smaller estimate for the Higgs contribution to the proton mass, around $9\%$ as given by the $\chi$QCD collaboration~\cite{Yang:2018nqn}.} This should be considered in modified gravity experiments that depend on the screening of the non-minimally coupled field, such as in atom interferometers~\cite{Burrage_2015,Elder_2016,AION}. 

The coupling of fifth forces to baryons has been previously considered, focusing on the nuclei of different elements~\cite{Damour:2010rm,Damour:2010rp}. However, it can be the case that, since quantum corrections are calculated once the theory is stable at its vacuum configuration (i.e., after calculating the vevs of the fields), they do not contribute to screening the fifth forces. For example, regarding the fifth force coupling to baryons, the main channel of the interaction comes from the conformal anomaly to the gauge field~\cite{Burrage:2018dvt} via the effective coupling to the would-be Higgs
\begin{equation}
    \lgr_{\rm eff}\propto \frac{\tilde{\phi}}{\langle\tilde{\phi}\rangle}G_{\mu\nu\,a}G^{\mu\nu\,a},
\end{equation}
where $\langle \tilde{\phi}\rangle=v_{\tilde{\phi}}$ and we have omitted the Wilson coefficient and proportional terms to the number of quarks producing this effective coupling. However, we can see that this term will not contribute to the field screening as it is normalised over the vev of the would-be Higgs field. Nonetheless, an actual computation of how bound states couple to the non-minimally coupled field is required before making a complete statement about the baryon contribution to screening mechanisms.

\section{Implications of screening mechanisms on fermion masses}\label{sec: over-screening}

Our analysis of screening mechanisms in the Standard Model has found that only those terms coupling to the Higgs will contribute to the screening of fifth forces. This has important implications, as particle masses experience a shift in their mass depending on the local value of the non-minimally coupled field's vev, which depends on the background density. 

Usually, screening effects are assumed to hide fifth forces without impacting how much a field is screened, naturally explaining the local lack of evidence for modified gravity while retaining $\tAv\approx1$. In this section, we will explore how valid this approximation is, as fine-tuning might be required to avoid fifth forces while not affecting the strength of gravitational interactions. We will do this using two popular models that present screening mechanisms: the chameleon~\cite{Burrage:2017qrf, Khoury:2003rn,Burrage_2016} and symmetron~\cite{Hinterbichler:2010es,Hinterbichler:2011ca}.
\begin{itemize}
    \item {\bf{Chameleon model:}} \edit{ These screening mechanisms suppress fifth forces by increasing the local mass of the non-minimally coupled field. While there are various formulations, for convenience, we will adopt the following one defined by the Jordan frame action} 
    \begin{equation}
        S=\int dx^4 \sqrt{-g}\left[-\frac{\alpha\varphi^2}{2}R +\frac{1}{2} g^{\mu\nu}\partial_\mu \varphi \partial_\nu \varphi - U(\varphi) + \lgr_{\rm m}\{\psi_i,g_{\mu\nu}\}\right],
    \end{equation}
    where $\alpha$ is a dimensionless constant and the potential has the generic power-law form\footnote{\edit{Other popular chameleon models define the power-law potential directly in the Einstein frame. Since these models exhibit similar runaway behavior, the main arguments presented in this section remain unchanged. Our choice, made for convenience, allows for an analytic expression of vev of the non-minimally coupled field as a function of the background's density.}}
    \begin{equation}
        U(\varphi)\propto \varphi^{4-n}.
    \end{equation}
    After taking the conformal transformation and canonically normalising the $\chi$ field into $\tilde{\chi}$ as demonstrated in section~\ref{sec:BD as BSM}, we obtain the coupling function
\begin{equation}
	\tilde{A}({\tilde{\chi}})=e^{\frac{\tilde{\chi}}{M}},
\end{equation}
and the potential
\begin{equation}
	\tilde{V}(\tilde{\chi})=V_0 e^{\frac{-n\tilde{\chi}}{M}},
\end{equation}
where $M$ is a generic mass scale related to the strength of fifth forces and $V_0$ is a constant term related to the mass of the scalar field. This potential is usually expanded into powers of $\tilde{\chi}$, but we will keep this generic form here for convenience. This type of action has also been widely studied in the context of the expansion of the universe as a popular quintessence model~\cite{Copeland:2006wr, Copeland:1997et}.

The next step is to obtain the vev of the $\tilde{\chi}$ field, for which we will need to use the screening equation from Eq.~\eqref{eq:vevs}, leading to
\begin{equation}
	\tilde{V}'(v_{\tilde{\chi}})=\frac{\tAvp{'}}{\tAv}\rho_\psi=\tAvp{'}yv\bar{\tilde{\psi}}\tilde{\psi}=\tAvp{'} \rho ,
\end{equation}
where \edit{$\rho_\psi=yv\tAv\bar{\tilde{\psi}}\tilde{\psi}$ is density as given by the mass calculated in Eq.~\eqref{eq: mass fermion} and }$\rho=yv\bar{\tilde{\psi}}\tilde{\psi}$ is the constant constituent of the density, equivalent to applying the locally measured value for the fermion masses to objects in different backgrounds.
With this, we find
\begin{equation}
	\frac{V_0 n}{M}e^{\frac{-nv_{\tilde{\chi}}}{M}}=\frac{1}{M} e^{\frac{v_{\tilde{\chi}}}{M}} \rho,
\end{equation}
leading to
\begin{equation}\label{eq: vev chi chameleon}
	\frac{v_{\tilde{\chi}}(\rho)}{M}=(1+n)\log\left(\frac{n \,V_0 }{\rho}\right),
\end{equation}
which is stable only for positive $n$. From this solution, we can see that although the exponent of $\tA$ is Planck suppressed, this cancels when substituting Eq.~\eqref{eq: vev chi chameleon}, leading to the following shift of elementary particles' masses
\begin{equation}
	m=yv\tAv= m_0 \left(\frac{n V_0 }{\rho}\right)^{1+n},
\end{equation}
where $m_0=yv$ is the particle's mass as measured locally. Therefore, big changes in the energy density of the background will introduce significant differences in the gravitational interaction of elementary particles. This effect can be larger than any possible fifth force, as can be seen by perturbing the couplings $\tA$ as in Eq.~\eqref{eq: A(chi+vev)}, giving
\begin{equation}
	\tilde{A}(\tilde{\chi}+v_{\tilde{\chi}})=\left(\frac{n V_0 }{\rho}\right)^{1+n}\left(1+\frac{\chi}{M} +\frac{\chi^2}{2M^2}+\dots\right).
\end{equation}
Thus, while the leading term (the one shifting the masses of the elementary particles) is not Planck suppressed, any direct coupling to the perturbations of this field will be Planck suppressed. Moreover, using our results from section~\ref{sec:BD as BSM}, in particular Eq.~\eqref{eq: meff}, the effective mass of the light mode is given by
\begin{equation}
	m_\sigma^2=V''(v_{\tilde{\chi}})+\frac{\tAvp{''}}{\tAv}\rho_\psi=V_ 0  \left(\frac{\rho}{n V_0 }\right)^{n(1+n)}\frac{n^2}{M^2}+\frac{\left(n V_0 \right)^{1+n}}{M^2\rho^{n}},
\end{equation}
which increases in high-density backgrounds, suppressing any long-range interaction transmitted by this field. Similarly, the fifth forces are also screened by the coupling itself to matter by
\begin{equation}
    \lgr\supset-yv\tAvp{'}\sigma\tilde{\bar{\psi}}\tilde{\psi}=-\frac{yv}{M}e^{v_{\tilde{\chi}}/M} \sigma\tilde{\bar{\psi}}\tilde{\psi}.
\end{equation}
Therefore, fifth forces completely disappear in high-density environments due to the increase of the propagator's effective mass and the decrease of their coupling to matter. However, this does not imply that all modifications from scalar-tensor theories vanish, as the mass shift of elementary particles can still lead to more important consequences, as we will explore in section~\ref{subsec:overscreening in galaxies}.

    \item {\bf{Symmetron model:}} It is defined by having the following non-minimally coupling function in the Einstein frame
\begin{equation}
    \tilde{A}(\tilde{\chi})=1+\frac{\tilde{\chi}^2}{2M_{\rm sym}^2},
\end{equation}
where $M_{\rm sym}$ is a mass scale related to the fifth force. Additionally, the symmetron potential takes the double-well form
\begin{equation}
    \tilde{V}(\tilde{\chi})=-\frac{1}{2}\mu_\chi^2 \tilde{\chi}^2+\frac{\lambda_\chi}{4!}\tilde{\chi}^4,
\end{equation}
which, when taking into consideration the local density background, leads to the effective potential
\begin{equation}
    V_{\rm eff}(\tilde{\chi})=\frac{1}{2}\left(\frac{\rho}{ M_{\rm sym}^2}-\mu_\chi^2\right)\tilde{\chi}^2+\frac{\lambda_\chi}{4!}\tilde{\chi}^4.
\end{equation}
Minimising this effective potential, we find that the vacuum expectation value for the field $\chi$ is given by
\begin{equation}
    v^2_{\tilde{\chi}}=\begin{cases}
      \frac{6}{\lambda_\chi}\left(\mu_\chi^2-\frac{\rho}{ M_{\rm sym}^2}\right) & \quad\text{for }{\rho}<\mu_\chi^2{ M_{\rm sym}^2}\\
      0 & \quad\text{for }{\rho}>\mu_\chi^2{ M_{\rm sym}^2}
    \end{cases}   
\end{equation}
where the field's $\mathbb{Z}_2$ symmetry is restored in high-density backgrounds, giving the name to the model. This will impact the effective mass of the fifth-force mediating mode by
\begin{align}
    m_{\sigma}^2=(2)\left|\frac{\rho}{ M_{\rm sym}^2}-\mu_\chi^2\right|,
\end{align}
where the factor of $(2)$ only multiplies when the field is unscreened (${\rho}<\mu_\chi^2{ M_{\rm sym}^2}$).  Moreover, the strength of the field will also be affected, leading to
\begin{equation}
    \lgr\supset - y v \tAvp{'} \sigma \bar{\tilde{\psi}} \tilde{\psi}= - y v \frac{v_{\tilde{\chi}}}{M_{\rm sym}^2} \sigma \bar{\tilde{\psi}} \tilde{\psi}.
\end{equation}
We can see that in high-density backgrounds, the field screens the fifth forces by increasing the mass of the propagator and decreasing the fifth force coupling. 
\end{itemize}

These two models are highly constrained due to their effect on the Weak Equivalence Principle and the impact on the dynamics of systems through fifth forces (see, for example, Refs.~\cite{Hinterbichler:2011ca,Khoury:2003rn,Burrage:2017qrf}). However, these bounds heavily depend on tests within the Solar System and the screening profile considered. For instance, it is usually assumed that the unscreened mass of the non-minimally coupled field is of the same order as the Universe's radius, such that it can explain the universe's expansion while producing fifth forces at large distances, which requires the usual fine-tuning related to the Cosmological Constant Problem~\cite{Weinberg:1988cp,Martin:2012bt}. On the contrary, having a larger mass in vacuum would evade most tests as fifth forces would be naturally exponentially suppressed~\cite{Hinterbichler:2011ca}. This is one of the main reasons why massive symmetron or chameleon fields \edit{are not commonly considered}, as they would lack phenomenological interest. Nevertheless, while fifth forces may not be effective for studying modified gravity within this region of the parameter space, studying the shift in fermion masses remains a viable option. This is because the masses would still be influenced by the screening of the field, resulting in a space-dependent gravitational force
\begin{equation}\label{eq:gravity}
    F_G=\frac{\tAvexp{2} m_0^2}{4\pi\tMpl^2 r^2},
\end{equation}
where $m_0$ is the mass of particles as measured locally, which can add up to make more massive, compact objects. As mentioned in section~\ref{sec:BD as BSM}, the gravitational force can be made a dimensionless observable by taking its ratio to the electric force.  Otherwise, we can take the ratio of gravitational interactions in different spacetime regions, corresponding to different values for $\tAv$. For example, for the chameleon model, we find the ratio
\begin{align}\label{eq: FG chameleon}
    \left(\frac{F_{G}}{F_{G,0}}\right)_{\rm cham}=\tAvexp{2} = \left(\frac{n V_0}{\rho}\right)^{2(n+1)},
    \end{align}
    where $F_{G,0}$ is the gravitational force in Eq.~\eqref{eq:gravity} for $\tAv=1$, corresponding to our local observations. This choice for the local value for $\tAv$ requires $nV_0\equiv \rho_0$, where $\rho_0$ is the density within the Solar System.\footnote{Here, one may opt to take the density of Earth instead, which may lead to a change of the strength of gravity in the vacuum around Earth. However, we will consider average densities since obtaining the exact space distribution of the non-minimally coupled field usually requires a numerical approach~\cite{2021MNRAS.506.5721A,Burrage:2023eol}.} In the symmetron case, we find that the local gravitational interaction depends on the background density by
    \begin{align}\label{eq: FG symmetron}
        \left(\frac{F_{G}}{F_{G,0}}\right)_{\rm sym}=\tAvexp{2} = \begin{cases}
      \left[1+\frac{6}{2\lambda_\chi M_{\rm sym}^2}\left(\mu_\chi^2 -\frac{\rho}{ M_{\rm sym}^2}\right)\right]^2 & \quad\text{for }{\rho}<\mu_\chi^2{ M_{\rm sym}^2}\\
      1 & \quad\text{for }{\rho}>\mu_\chi^2{ M_{\rm sym}^2}
    \end{cases}    
\end{align}
where allowing for any value for $\mu_\chi$ can lead to big deviations on $\tAv$. Expecting $\tAv=1$ locally implies that the symmetron must be screened within the Solar System, such that $\rho_0>\mu_\chi^2M_{\rm sym}^2$.

Note that screening mechanisms will always tend to monotonically decrease the mass of elementary particles in high-density environments,\footnote{ A short proof for this statement: In very high densities, the effective potential $V_{\rm eff}(\tilde{\chi})=\tilde{V}(\tilde{\chi})+\tA \rho$ is dominated by the coupling function $\tA$, so $v_{\tilde{\chi}}$ must be located at a local minimum of $\tAv$. Therefore, the shifted mass, $m=m_0\tAv$, takes its minimum value at high densities, as any shift on the effective minimum of $V_{\rm eff}$ given by the dependence on $V$ will always shift the field to a higher value for $\tAv$. This means that the lower the density, the higher the mass for elementary particles.} but this rate of change will be model-dependent. For example, the chameleon model presents \textit{over-screening}, as the shift in the masses of elementary particles is very sensitive to changes in the local density, even when fifth forces are completely suppressed. On the other hand, for the symmetron model, the mass of particles is always constant once the fifth forces are screened, for densities higher than $\rho_c=\mu_\chi^2{ M_{\rm sym} ^2}$. 

\subsection{Over-screening effects on galactic disks}\label{subsec:overscreening in galaxies}
The evolution of the mass for Standard Model particles has already been considered regarding the expansion of the universe, as the scalar field's roll under the influence of its own potential will give a time-varying mass to the elementary particles. For example, this has been applied from neutrino physics, addressing their small mass~\cite{Brookfield:2005bz} or the Hubble tension~\cite{Sakstein:2019fmf}, up to inflation in the context of string theory~\cite{Cicoli:2023opf,Casas:2024xqy}. However, here, we relate this mass change to screening mechanisms and their effect on elementary particles in late-time cosmology, where the non-minimally coupled field is already stable at its effective minimum. Therefore, instead of studying the time evolution of the scalar field, we will focus on the implications of the distribution of matter at current times. As an illustration, we will see the impact of screening mechanisms on galactic disks.

For this, we will use the following density function for the galactic disk
\begin{equation}\label{eq: rho disk}
    \rho_{\rm disk,0}(r,z)=\frac{\Sigma_{\rm disk}}{2z_{\rm disk}} e^{-r/r_{\rm disk}}e^{-|z|/z_{\rm disk}},
\end{equation}
where this system is in cylindrical coordinates, and $\Sigma_{\rm disk},~z_{\rm disk}$ and $r_{\rm disk}$ are length scales characterizing different galaxies. Following Ref.~\cite{Burrage:2023eol}, these quantities can be reparametrised into the virial mass, $M_{\rm vir}$ (defined as the enclosed mass within the radius at which the galaxy density equals 200 times the critical density of the universe). Taking $M_{\rm vir}=10^{11.5} M_\odot$ corresponds to $\Sigma_{\rm disk}=6.4\times 10^8 M_\odot \text{kpc}^{-2},~z_{\rm disk}=0.2 \text{kpc}$ and $r_{\rm disk}=0.6 \text{kpc}$. For a realistic illustration of this scenario, we would also have to consider other galaxy components, such as dark matter\footnote{Dark matter will also feel the mass shift due to screening mechanisms, unless it obtains its mass by a dynamically broken scale symmetry, as we saw in section~\ref{subsec SM}.} and the bulge~\cite{2017MNRAS.465...76M}, but we will neglect them in this illustration for simplicity. Similarly, we will take that baryonic matter contributes with a $100\%$ of its mass, as any other percentage can be redefined into the model parameters.

\begin{figure}
    \centering
    \includegraphics[trim={2cm 0 2cm 0},clip,width=1\textwidth]{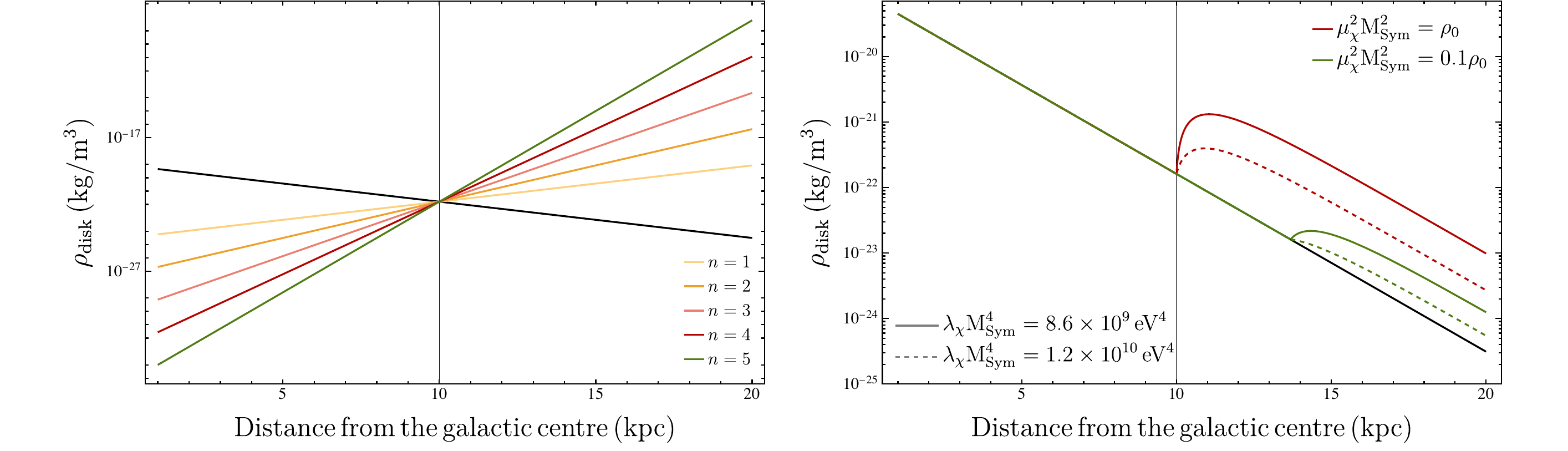}
    \caption{Impact of screening mechanisms on the density of a galactic disk as given in Eq.~\eqref{eq: rho disk} (black line) for the chameleon (left) and symmetron (right) models. Both models have been tuned to give $\tAv=1$ in the Solar System, at $10$kpc from the galactic center. We can see that the chameleon model presents over-screening, while the symmetron's rescaling of the masses is constant after reaching the critical density $\rho_c=\mu_\chi^2 M_{\rm Sym}^2$. }
    \label{fig: densities}
\end{figure}

We calculate the \edit{effective rescaled density by evaluating $\tAv$ at each point through}
\begin{equation}
    \rho_{\rm disk}(r,z)=\rho_{\rm disk,0}(r,z) \tilde{A}(v_{\tilde{\chi}}(\rho_{\rm disk,0}(r,z)).
\end{equation}
\edit{For this, we have assumed that the matter density in the galaxy disk is measured using Earth's values for the observed particle masses. To express these values to their true, rescaled masses, we have applied Eq. \eqref{eq: mass fermion}. It is important to emphasise that this is a toy model calculation for illustrative purposes. A more realistic scenario would need to account for the mass shift's impact on the stars' luminosity, which would change our current estimate for Eq.~\eqref{eq: rho disk}. Moreover, for a more accurate estimation of this effect within galaxies, it is crucial to compute the \textit{screening map}, as done in Refs.\cite{2021MNRAS.506.5721A,Burrage:2023eol}, or to use numerical simulations, such as those in Refs.~\cite{Ruan:2021wup, Hernandez-Aguayo:2021kuh}.}

We present the results in Figure~\ref{fig: densities}, with one plot for each model, where the black line is the density distribution of the galactic disk as in Eq.~\eqref{eq: rho disk} evaluated at $z=0$, and the rest of the lines correspond to the real, shifted mass of the galactic disk for different models and sets of parameters. On the left, we can see that the chameleon model is very sensitive to the background density, needing fine-tuning to generate the observed local strength of gravity, using $nV_0=\rho_0$ (Eq.~\eqref{eq: vev chi chameleon})\footnote{\edit{Had we chosen the usual negative power-law potential defined in the Einstein frame for the Chameleon field, the effect on fermion masses would be less noticeable at high densities. This is because such potentials asymptote at $\chi=0$, setting a minimum possible value for $\tAv$. However, given the runaway potential, we would expect the same tendency at low-density backgrounds as in Figure~\ref{fig: densities}.}}. On the right, the symmetron model shows that a constant strength of gravity is obtained inside the critical density $\rho_c=\mu_\chi^2 M_{\rm sym}^2$, where all fifth forces are suppressed. In this case, changing $\mu_\chi M_{\rm sym}$ shifts the critical mass, while changing $\lambda_\chi M_{\rm sym}^4$  impacts the mass rescaling outside the screened region.


Thus, even though we commonly rely on screening mechanisms to eliminate the effects introduced by modified gravity due to fifth forces, these mechanisms can lead to more significant consequences, such as the mass shift of elementary particles. Although we neglected the contribution of fifth forces in this section, there might be regions where both effects, those the mass shift and long-range forces, become relevant.

\section{Conclusion}\label{sec: conclusion}

Screening mechanisms are a popular method of evading the tight constraints of fifth forces in scalar-tensor theories. These rely on the non-minimal coupling of the additional scalar singlet to the Ricci scalar, relating the strength of the introduced long-range forces to the local curvature, which is connected to the local density. 

Since these theories include an extra scalar in the action, they can also be studied from a particle physics perspective. This has been done in vacuum, focusing on the emergence of long-range fifth forces and additional interactions in the Standard Model~\cite{Burrage:2018dvt,Copeland:2021qby}. This paper has expanded previous work in Ref.~\cite{Burrage:2023eol} to account for screening effects in this context, with the main difference being that we relaxed the assumption of $\tAv\approx1$, where $\tAv$ is the late time value for the non-minimal coupling function. We showed that in the Standard Model, the Higgs field does not alter the effective potential for the non-minimally coupled field, obtaining the same result as in the standard treatment of screening mechanisms. We further explored how different mass-generating mechanisms contributed to the screening of fifth forces. In particular, we first studied explicit mass terms and scale invariant theories that generate their mass dynamically. In the former case, we find that the contribution is equal to that of the Higgs mechanism, while in the latter, the field does not contribute. Therefore, there is a relation between the fifth force and the contribution to the screening, as expected. Considering this, we found that baryonic matter would only contribute via the percentage that gains its mass via the Higgs mechanism, between $10\%$ and $20\%$ for the mass of the proton, depending on the method used to study the proton's mass.

\edit{In the Einstein frame, the coupling of the non-minimally coupled field to the Higgs leads to a mixing in their vacuum expectation values. This has important consequences on the Standard Model,} as screening mechanisms shift the mass of elementary particles, as discussed after Eq.~\eqref{eq: mass fermion}. This effect can be significant even when fifth forces are completely suppressed, meaning that, for some models, a field cannot be infinitely screened without consequences on the physics, a phenomenon we have called \textit{over-screening}. We studied this effect on \edit{ a toy model for} galactic disks for the chameleon and symmetron models, finding that changes in the local density give a different gravitational potential to the one expected by standard gravity, as shown in Figure~\ref{fig: densities}. In particular, the chameleon model \edit{ with an exponential potential} is a perfect example of over-screening, as independently of its mass and coupling strength to matter, it will drastically change all fermion masses depending on the background's density, as given by Eq.~\eqref{eq: FG chameleon}. On the other hand, the symmetron model leads to a constant mass inside the screened region (i.e., for $\rho>\mu_\chi^2 M^2_{\rm sym}$). Outside this region, the masses rapidly converges to a rescaling given by $m_{\psi}/m_0=(1+\mu^2/(2\lambda_\chi M_{\rm sym}^2))$, as we show in Eq.~\eqref{eq: FG symmetron}. Thus, increasing the mass of the symmetron field leads to more significant effects, but it can always be compensated by increasing $\lambda_\chi$ or $M_{\rm sym}$.

The spatially dependent \edit{fermion masses} is a new window for phenomenology in modified theories of gravity in a region of the parameter space that previously seemed uninteresting. For example, this could be a great contender for dark matter as it increases the gravitational potential in low-density environments. However, the main challenge lies in the fact that \edit{our understanding of galactic dynamics comes from the nuclear processes taking place inside stars, which would be modified by the shift in fermion masses depending on the model.} In future work, we aim to better study the behaviour of matter for these models, such as in the stability of galaxies or as a possible explanation for the observed radial acceleration relations~\cite{2016PhRvL.117t1101M}, following the work from Ref.~\cite{Burrage:2016yjm}. Additionally, over-screening can also have strong implications for quintessence scenarios motivated by scalar-tensor theories, such as the exponential potential in the chameleon model~\cite{Copeland:2006wr, Copeland:1997et}, where the coupling to matter would introduce non-linear effects to the evolution of the fields. Furthermore, since scalar-tensor theories are related to the low energy description for string theory, we can also relate the over-screening of a field to the Swampland conjectures~\cite{Grana:2021zvf,Vafa:2005ui}. For example, the distance conjecture would not only restrict $\Delta v_{\tilde{\chi}}\ll\tMpl$ during the evolution of the field in time, but also in the late-time distribution of energy density.

There are many positive aspects of studying modified gravity from a particle perspective, such as consistently studying the modification of gravitational interactions along those in the Standard Model. Moreover, throughout this work, we have only considered the classical aspects of the scalar-tensor theories, but this description will allow us to account for quantum corrections, which is a topic worthy of further exploration. 

\section*{Acknowledgements}
We thank Clare Burrage, \edit{ Clifford Burgess}, Edmund Copeland, Baojiu Li, Iv\'an Mart\'inez Soler, Matteo Marcoli, Peter Millington, \edit{ and Javier Rubio} for useful discussions. This work is supported by the STFC under Grant No.~ST/T001011/1.

\appendix 
\section{Screening effects in the Jordan frame}\label{Appendix A}
In section \ref{subsec SM}, we found that when a fermionic field obtains its mass through a dynamical symmetry breaking, it does not contribute to the screening of the minimally coupled field. Moreover, we also found that this field wouldn't experience the mass shift found for the explicit symmetry-breaking case. As explained in section~\ref{sec:BD as BSM}, the mass shift we find in the Einstein frame comes from interactions with the Higgs field. In the Jordan frame, these effects arise as a shift in the Planck mass, as we will show now. 

We will use the combined scale breaking action from section~\ref{subsec SM},
\begin{align}
	S[g_{\mu \nu}]=\int \D^4{x} \sqrt{-g}& \left[-\frac{F(\varphi)}{2}R+\frac{1}{2}g^{\mu \nu}\partial_\mu \phi \partial_\nu \phi +\frac{1}{2}g^{\mu \nu}\partial_\mu \theta \partial_\nu \theta +\frac{1}{2}g^{\mu\nu}\partial_\mu\varphi\partial_\nu\varphi\right.\nonumber\\
	&\quad+\bar{\psi}\gamma^\mu \nabla_\mu\psi- y\bar{\psi}\phi\psi +\bar{\psi} \gamma^\mu \nabla_\mu\psi- m\bar{\psi}\psi-U(\varphi)\nonumber\\ &\quad \left.-W(\phi, \theta)+\frac{1}{2} \mu_{\theta}^{2} A^{-2}(\varphi)\theta^{2}-\frac{\lambda_{\theta}}{4 !} \theta^{4}-\frac{3}{2} \frac{\mu_{\theta}^{4}}{\lambda_{\theta}} A^{-4}(\varphi)\right],
	\label{sm app}
\end{align}
where we remind the reader that $F(\varphi)=\tMpl^2 A^{-2}(\varphi)$. The potential is given by
\begin{equation}
	W(\phi, \theta)=\frac{\lambda}{4 !}\left(\phi^{2}-\frac{\beta}{\lambda} \theta^{2}\right)^{2}-\frac{1}{2} \mu^{2}\left(\phi^{2}-\frac{\beta}{\lambda} \theta^{2}\right)+\frac{3}{2} \frac{\mu^{4}}{\lambda},\label{comb_pot app}
\end{equation}
where we have two extreme cases, the explicitly broken scale symmetry limit ($\beta\to0$) and the dynamically broken limit ($\mu\to0$).

In the Jordan frame, the screening of the non-minimally coupled field does not come from the interaction with the Higgs, but through the non-minimal coupling to the Ricci scalar. In this way, the equation of motion for the $\varphi$ field leads to
\begin{equation}\label{eq: eomvarphi}
	\square \varphi +U'(\varphi)+\frac{F'(\varphi)}{2}R=0,
\end{equation}	
where the value for the Ricci scalar can be obtained from the modified Einstein equation
\begin{equation}\label{eq: Einsten Jordan}
	G_{\mu\nu}=\frac{1}{F(\varphi)}\left(T_{\mu\nu}^{\rm (m)}+T_{\mu\nu}^{(d)}+T_{\mu\nu}^{(\varphi)}\right),
\end{equation}
where $T_{\mu\nu}^{\rm (m)}$ is the energy-momentum tensor of the matter sector and $T_{\mu\nu}^{\varphi}$ of the non-minimally coupled field. Taking the trace of the Einstein equation and using ${T^{\rm (i)}}_{\mu}^\mu=\rho_i$, with $\rho_m\gg\rho_\varphi$, we obtain
\begin{equation}
	R=-\frac{1}{F(\varphi)}\rho_{m},
\end{equation}
where the matter sourced by the dynamical scale breaking won't contribute to $\rho_m$ since the trace of their energy-momentum tensor vanishes~\cite{Ferreira:2016kxi}. Substituting this equation into Eq.~\eqref{eq: eomvarphi}, we find
\begin{equation}
	\square \varphi +U'(\varphi)-\frac{F'(\varphi)}{2F(\varphi)}\rho_m=0,
\end{equation}	
where, using $F(\varphi)=\tMpl^2/A^2(\varphi)$, leads to the Jordan frame effective potential
\begin{equation}\label{eq: Veff app}
	V_{\rm eff}(\varphi)=U(\varphi) +\log(A(\varphi))\rho_m,
\end{equation}
 equivalent to the Einstein frame effective potential in Eq.~\eqref{eq:vevs}. We will now study the implications of the screening of the non-minimally coupled field on the Standard Model in both limit cases:
\subsubsection*{Explicitly broken limit ($\beta\to0$)}
Taking this limit, the extra field $\theta$ decouples and we recover the usual double-well potential for the would-be Higgs sector, as in section~\ref{sec:BD as BSM}. As we can see, the masses do not shift with the screening mechanism since there are no extra couplings of the non-minimally coupled field to the Higgs\edit{. However, the propagation of particles will still be affected by the modified geodesic equations in the Jordan frame (which can be derived from Eq.~\eqref{eq: Einsten Jordan})}. Similarly, all gravitational interactions will get modified because of the definition of the effective Planck mass, defined via
\begin{equation}
	\Mpl^2(v_\varphi)\equiv F(v_\varphi),
\end{equation}
which is now environmentally dependent via the effective potential in Eq.~\eqref{eq: Veff app}. We can see that this leads to the same result as in the Einstein frame by comparing dimensionless quantities, such as the ratio of gravity against the electric force, given in the Jordan frame by
\begin{equation}
	\left(\frac{F_{\rm G}}{F_{\rm EM}}\right)_{\rm JF}=\frac{\frac{1}{4\pi r^2} \frac{m^2}{F(v_\varphi)}}{\frac{1}{4\pi r^2} {q^2}}=\frac{m^2 }{q^2F(v_\varphi)},
\end{equation}
where $q$ is the charge of the test particles. The same calculation done in the Einstein frame leads to
\begin{equation}
	\left(\frac{F_{\rm G}}{F_{\rm EM}}\right)_{\rm EF}=\frac{\frac{1}{4\pi r^2} \frac{m^2 \tAvexp{2}}{\tMpl^2}}{\frac{1}{4\pi r^2} q^2}=\frac{m^2\tAvexp{2}}{q^2\tMpl^2},
\end{equation}
where we have introduced the mass shifted mass $m \tAv$ and the Einstein frame's constant Planck mass  $\tMpl$. We thus find the equivalence between frames by considering the definition $\tAvexp{2}=\tMpl^2/F(\varphi)$. 

\subsubsection*{Dynamically broken limit ($\mu\to 0$)}
While the fermions will not contribute to screening the fifth forces in this limit, we will consider that a different sector is driving the screening effects. This allows us to see the impact screening would have on a scale-invariant sector by studying the mass of the fermions. For this, we have to find the minimum of the scalar field potential from Eq.~\eqref{sm app}, which is given by solving the following system of equations
\begin{equation}
	\begin{split}
	&\frac{\lambda}{6}v_{{\phi}}^4-\frac{\beta}{6}v_{{\theta}}^2v_{{\phi}}^2=0,\\
	&\frac{\beta^2}{6\lambda}v_{{\theta}}^4-\frac{\beta}{6}v_{{\theta}}^2v_{{\phi}}^2-\mu_\theta^2A^{-2}(v_\varphi) v_{{\theta}}^2-\frac{\lambda_\theta}{6}v_{{\theta}}^4=0\\ 
	&U'(v_\varphi)-\frac{F'(v_\varphi)}{F(v_\varphi)}\rho_m-\mu_\theta^2v_\theta ^2A^{-3}(v_\varphi)A'(v_\varphi)+6\frac{\mu_\theta^4}{\lambda_\theta}A^{-5}(v_\varphi)A'(v_\varphi)=0.
\end{split}
\end{equation}
For any generic ${A}({\varphi})$ and ${U}({\varphi})$, it gives
\begin{align}
&v_{{\phi}}^2=\frac{\beta}{\lambda}v_{{\theta}}^2\\
& v_{{\theta}}^2=\frac{6\mu_\theta^2}{\lambda_\theta}A^{-2}(v_\varphi),\\
&{U}'(v_{{\varphi}})=\frac{F'(v_\varphi)}{F(v_\varphi)}\rho_m,
\end{align}
agreeing with the Einstein frame result. Expanding the scalar fields around their vacuum expectation values, we find the following mass for the fermionic fields
\begin{equation}
	m=y v_\phi=yA^{-1}(v_\varphi)\sqrt{\frac{\beta}{\lambda}\frac{6\mu_\theta^2}{\lambda_\theta}},
\end{equation}
which is environmentally dependent through the screening of $\varphi$, contrary to what we found in the Einstein frame. \edit{However, this rescaling perfectly cancels the modification in the geodesic equation, recovering the standard propagation in Einstein's gravity for all particles}. 

This result agrees with the one found in the Einstein frame, as can be seen by comparing dimensionless observables. Taking the ratio of the gravitational and electric forces for the pure dynamical scale breaking case in the Jordan frame, we find
\begin{equation}
	\left(\frac{F_{\rm G}}{F_{\rm EM}}\right)_{\rm JF}=\frac{\frac{1}{4\pi r^2} \frac{m^2}{F(v_\varphi)}}{\frac{1}{4\pi r^2} {q^2}}=\frac{yA^{-2}(\varphi)\frac{\beta}{\lambda}\frac{6\mu_\theta^2}{\lambda_\theta} }{q^2F(v_\varphi)}.
\end{equation}
Using the definition of $A^{2}(\varphi)=\tMpl^2/F(\varphi)$, we obtain the exact same result as in the Einstein frame,
\begin{equation}
	\left(\frac{F_{\rm G}}{F_{\rm EM}}\right)_{\rm EF}=\frac{y\frac{\beta}{\lambda}\frac{6\mu_\theta^2}{\lambda_\theta}}{q^2\tMpl^2}.
 \end{equation}

Therefore, we see that screening the non-minimally coupled field always tends to change the local Planck mass value, affecting the strength of all gravity-related forces. \edit{However, if the matter sector generates the masses by a dynamical scale symmetry breaking, all fields will be invariant to such changes.}
\bibliographystyle{IEEEtran}
\bibliography{mybibs}

\begin{thebibliography}{10}
\providecommand{\url}[1]{#1}
\csname url@samestyle\endcsname
\providecommand{\newblock}{\relax}
\providecommand{\bibinfo}[2]{#2}
\providecommand{\BIBentrySTDinterwordspacing}{\spaceskip=0pt\relax}
\providecommand{\BIBentryALTinterwordstretchfactor}{4}
\providecommand{\BIBentryALTinterwordspacing}{\spaceskip=\fontdimen2\font plus
\BIBentryALTinterwordstretchfactor\fontdimen3\font minus \fontdimen4\font\relax}
\providecommand{\BIBforeignlanguage}[2]{{%
\expandafter\ifx\csname l@#1\endcsname\relax
\typeout{** WARNING: IEEEtran.bst: No hyphenation pattern has been}%
\typeout{** loaded for the language `#1'. Using the pattern for}%
\typeout{** the default language instead.}%
\else
\language=\csname l@#1\endcsname
\fi
#2}}
\providecommand{\BIBdecl}{\relax}
\BIBdecl

\bibitem{Yasunori:Fujii_2003}
Y.~Fujii and K.~ichi Maeda, ``The scalar-tensor theory of gravitation,'' \emph{Classical and Quantum Gravity}, vol.~20, no.~20, p. 4503, oct 2003.

\bibitem{Herranen:2014cua}
M.~Herranen, T.~Markkanen, S.~Nurmi, and A.~Rajantie, ``{Spacetime curvature and the Higgs stability during inflation},'' \emph{Phys. Rev. Lett.}, vol. 113, no.~21, p. 211102, 2014.

\bibitem{Markkanen:2018bfx}
T.~Markkanen, S.~Nurmi, A.~Rajantie, and S.~Stopyra, ``{The 1-loop effective potential for the Standard Model in curved spacetime},'' \emph{JHEP}, vol.~06, p. 040, 2018.

\bibitem{Steinwachs:2011zs}
C.~F. Steinwachs and A.~Y. Kamenshchik, ``{One-loop divergences for gravity non-minimally coupled to a multiplet of scalar fields: calculation in the Jordan frame. I. The main results},'' \emph{Phys. Rev. D}, vol.~84, p. 024026, 2011.

\bibitem{Cicoli:2023opf}
M.~Cicoli, J.~P. Conlon, A.~Maharana, S.~Parameswaran, F.~Quevedo, and I.~Zavala, ``{String Cosmology: from the Early Universe to Today},'' 3 2023.

\bibitem{Horndeski:1974wa}
G.~W. Horndeski, ``{Second-order scalar-tensor field equations in a four-dimensional space},'' \emph{Int. J. Theor. Phys.}, vol.~10, pp. 363--384, 1974.

\bibitem{Kobayashi:2019hrl}
T.~Kobayashi, ``{Horndeski theory and beyond: a review},'' \emph{Rept. Prog. Phys.}, vol.~82, no.~8, p. 086901, 2019.

\bibitem{Traykova:2019oyx}
D.~Traykova, E.~Bellini, and P.~G. Ferreira, ``{The phenomenology of beyond Horndeski gravity},'' \emph{JCAP}, vol.~08, p. 035, 2019.

\bibitem{Gleyzes:2013ooa}
J.~Gleyzes, D.~Langlois, F.~Piazza, and F.~Vernizzi, ``{Essential Building Blocks of Dark Energy},'' \emph{JCAP}, vol.~08, p. 025, 2013.

\bibitem{Langlois:2015cwa}
D.~Langlois and K.~Noui, ``{Degenerate higher derivative theories beyond Horndeski: evading the Ostrogradski instability},'' \emph{JCAP}, vol.~02, p. 034, 2016.

\bibitem{Langlois:2018dxi}
D.~Langlois, ``{Dark energy and modified gravity in degenerate higher-order scalar extendash{}tensor (DHOST) theories: A review},'' \emph{Int. J. Mod. Phys. D}, vol.~28, no.~05, p. 1942006, 2019.

\bibitem{Brans:1961sx}
C.~Brans and R.~H. Dicke, ``{Mach's principle and a relativistic theory of gravitation},'' \emph{Phys. Rev.}, vol. 124, pp. 925--935, 1961.

\bibitem{Wetterich:1987fm}
C.~Wetterich, ``{Cosmology and the Fate of Dilatation Symmetry},'' \emph{Nucl. Phys. B}, vol. 302, pp. 668--696, 1988.

\bibitem{Buchmuller:1988cj}
W.~Buchmuller and N.~Dragon, ``{Dilatons in Flat and Curved Space-time},'' \emph{Nucl. Phys. B}, vol. 321, pp. 207--231, 1989.

\bibitem{Shaposhnikov:2008xb}
M.~Shaposhnikov and D.~Zenhausern, ``{Scale invariance, unimodular gravity and dark energy},'' \emph{Phys. Lett. B}, vol. 671, pp. 187--192, 2009.

\bibitem{Shaposhnikov:2008xi}
------, ``{Quantum scale invariance, cosmological constant and hierarchy problem},'' \emph{Phys. Lett. B}, vol. 671, pp. 162--166, 2009.

\bibitem{Blas:2011ac}
D.~Blas, M.~Shaposhnikov, and D.~Zenhausern, ``{Scale-invariant alternatives to general relativity},'' \emph{Phys. Rev. D}, vol.~84, p. 044001, 2011.

\bibitem{Garcia-Bellido:2011kqb}
J.~Garcia-Bellido, J.~Rubio, M.~Shaposhnikov, and D.~Zenhausern, ``{Higgs-Dilaton Cosmology: From the Early to the Late Universe},'' \emph{Phys. Rev. D}, vol.~84, p. 123504, 2011.

\bibitem{Garcia-Bellido:2012npk}
J.~Garcia-Bellido, J.~Rubio, and M.~Shaposhnikov, ``{Higgs-Dilaton cosmology: Are there extra relativistic species?}'' \emph{Phys. Lett. B}, vol. 718, pp. 507--511, 2012.

\bibitem{Bezrukov:2012hx}
F.~Bezrukov, G.~K. Karananas, J.~Rubio, and M.~Shaposhnikov, ``{Higgs-Dilaton Cosmology: an effective field theory approach},'' \emph{Phys. Rev. D}, vol.~87, no.~9, p. 096001, 2013.

\bibitem{Henz:2013oxa}
T.~Henz, J.~M. Pawlowski, A.~Rodigast, and C.~Wetterich, ``{Dilaton Quantum Gravity},'' \emph{Phys. Lett. B}, vol. 727, pp. 298--302, 2013.

\bibitem{Rubio:2014wta}
J.~Rubio and M.~Shaposhnikov, ``{Higgs-Dilaton cosmology: Universality versus criticality},'' \emph{Phys. Rev. D}, vol.~90, p. 027307, 2014.

\bibitem{Karananas:2016grc}
G.~K. Karananas and M.~Shaposhnikov, ``{Scale invariant alternatives to general relativity. II. Dilaton properties},'' \emph{Phys. Rev. D}, vol.~93, no.~8, p. 084052, 2016.

\bibitem{Ferreira:2016vsc}
P.~G. Ferreira, C.~T. Hill, and G.~G. Ross, ``{Scale-Independent Inflation and Hierarchy Generation},'' \emph{Phys. Lett. B}, vol. 763, pp. 174--178, 2016.

\bibitem{Ferreira:2016kxi}
------, ``{No fifth force in a scale invariant universe},'' \emph{Phys. Rev. D}, vol.~95, no.~6, p. 064038, 2017.

\bibitem{Casas:2017wjh}
S.~Casas, M.~Pauly, and J.~Rubio, ``{Higgs-dilaton cosmology: An inflation extendash{}dark-energy connection and forecasts for future galaxy surveys},'' \emph{Phys. Rev. D}, vol.~97, no.~4, p. 043520, 2018.

\bibitem{Ferreira:2018qss}
P.~G. Ferreira, C.~T. Hill, J.~Noller, and G.~G. Ross, ``{Inflation in a scale invariant universe},'' \emph{Phys. Rev. D}, vol.~97, no.~12, p. 123516, 2018.

\bibitem{Damour:1992we}
T.~Damour and G.~Esposito-Farese, ``{Tensor multiscalar theories of gravitation},'' \emph{Class. Quant. Grav.}, vol.~9, pp. 2093--2176, 1992.

\bibitem{doneva2022scalarization}
D.~D. Doneva, F.~M. Ramazanoğlu, H.~O. Silva, T.~P. Sotiriou, and S.~S. Yazadjiev, ``Scalarization,'' 2022.

\bibitem{Cardoso:2013opa}
V.~Cardoso, I.~P. Carucci, P.~Pani, and T.~P. Sotiriou, ``{Matter around Kerr black holes in scalar-tensor theories: scalarization and superradiant instability},'' \emph{Phys. Rev. D}, vol.~88, p. 044056, 2013.

\bibitem{Avilez:2013dxa}
A.~Avilez and C.~Skordis, ``{Cosmological constraints on Brans-Dicke theory},'' \emph{Phys. Rev. Lett.}, vol. 113, no.~1, p. 011101, 2014.

\bibitem{Bertotti:2003rm}
B.~Bertotti, L.~Iess, and P.~Tortora, ``{A test of general relativity using radio links with the Cassini spacecraft},'' \emph{Nature}, vol. 425, pp. 374--376, 2003.

\bibitem{Fischer:2024eic}
H.~Fischer, C.~K\"ading, and M.~Pitschmann, ``{Screened Scalar Fields in the Laboratory and the Solar System},'' \emph{Universe}, vol.~10, p. 297, 2024.

\bibitem{EotWash}
S.~Bae\ss{}ler, B.~R. Heckel, E.~G. Adelberger, J.~H. Gundlach, U.~Schmidt, and H.~E. Swanson, ``Improved test of the equivalence principle for gravitational self-energy,'' \emph{Phys. Rev. Lett.}, vol.~83, pp. 3585--3588, Nov 1999.

\bibitem{Merkowitz:2010kka}
S.~M. Merkowitz, ``{Tests of Gravity Using Lunar Laser Ranging},'' \emph{Living Rev. Rel.}, vol.~13, p.~7, 2010.

\bibitem{Burrage:2017qrf}
C.~Burrage and J.~Sakstein, ``{Tests of Chameleon Gravity},'' \emph{Living Rev. Rel.}, vol.~21, no.~1, p.~1, 2018.

\bibitem{Williams:2005rv}
J.~G. Williams, S.~G. Turyshev, and D.~H. Boggs, ``{Lunar laser ranging tests of the equivalence principle with the earth and moon},'' \emph{Int. J. Mod. Phys. D}, vol.~18, pp. 1129--1175, 2009.

\bibitem{Khoury:2003rn}
J.~Khoury and A.~Weltman, ``{Chameleon cosmology},'' \emph{Phys. Rev. D}, vol.~69, p. 044026, 2004.

\bibitem{Burrage_2016}
C.~Burrage and J.~Sakstein, ``A compendium of chameleon constraints,'' \emph{Journal of Cosmology and Astroparticle Physics}, vol. 2016, no.~11, pp. 045--045, nov 2016.

\bibitem{Hinterbichler:2010es}
K.~Hinterbichler and J.~Khoury, ``{Symmetron Fields: Screening Long-Range Forces Through Local Symmetry Restoration},'' \emph{Phys. Rev. Lett.}, vol. 104, p. 231301, 2010.

\bibitem{Hinterbichler:2011ca}
K.~Hinterbichler, J.~Khoury, A.~Levy, and A.~Matas, ``{Symmetron Cosmology},'' \emph{Phys. Rev. D}, vol.~84, p. 103521, 2011.

\bibitem{deRham:2021fpu}
C.~de~Rham, S.~Melville, and J.~Noller, ``{Positivity bounds on dark energy: when matter matters},'' \emph{JCAP}, vol.~08, p. 018, 2021.

\bibitem{Brax:2022olf}
P.~Brax, A.-C. Davis, and B.~Elder, ``{Screened scalar fields in hydrogen and muonium},'' \emph{Phys. Rev. D}, vol. 107, no.~4, p. 044008, 2023.

\bibitem{Burrage:2018pyg}
C.~Burrage, C.~K\"ading, P.~Millington, and J.~Min\'a\v{r}, ``{Open quantum dynamics induced by light scalar fields},'' \emph{Phys. Rev. D}, vol. 100, no.~7, p. 076003, 2019.

\bibitem{Kading:2023mdk}
C.~K\"ading, M.~Pitschmann, and C.~Voith, ``{Dilaton-induced open quantum dynamics},'' \emph{Eur. Phys. J. C}, vol.~83, no.~8, p. 767, 2023.

\bibitem{Brax_2009}
P.~Brax, C.~Burrage, A.-C. Davis, D.~Seery, and A.~Weltman, ``Collider constraints on interactions of dark energy with the standard model,'' \emph{Journal of High Energy Physics}, vol. 2009, no.~09, pp. 128--128, sep 2009.

\bibitem{Argyropoulos:2023pmy}
S.~Argyropoulos, C.~Burrage, and C.~Englert, ``{Environmentally aware displaced vertices},'' 4 2023.

\bibitem{SevillanoMunoz:2023loj}
S.~Sevillano Mu\~noz, ``{FeynMG: Automating particle physics calculations in scalar-tensor theories},'' Ph.D. dissertation, Nottingham U., 2023.

\bibitem{Burrage:2018dvt}
C.~Burrage, E.~J. Copeland, P.~Millington, and M.~Spannowsky, ``{Fifth forces, Higgs portals and broken scale invariance},'' \emph{JCAP}, vol.~11, p. 036, 2018.

\bibitem{Copeland:2021qby}
E.~J. Copeland, P.~Millington, and S.~S. Mu\~noz, ``{Fifth forces and broken scale symmetries in the Jordan frame},'' \emph{JCAP}, vol.~02, no.~02, p. 016, 2022.

\bibitem{Burrage:2023eol}
C.~Burrage, B.~March, and A.~P. Naik, ``{Accurate computation of the screening of scalar fifth forces in galaxies},'' \emph{JCAP}, vol.~04, p. 004, 2024.

\bibitem{Williams_2004}
J.~G. Williams, S.~G. Turyshev, and D.~H. Boggs, ``Progress in lunar laser ranging tests of relativistic gravity,'' \emph{Phys. Rev. Lett.}, vol.~93, p. 261101, Dec 2004.

\bibitem{DeFelice:2010aj}
A.~De~Felice and S.~Tsujikawa, ``{f(R) theories},'' \emph{Living Rev. Rel.}, vol.~13, p.~3, 2010.

\bibitem{Muller:2007zzb}
J.~Muller and L.~Biskupek, ``{Variations of the gravitational constant from lunar laser ranging data},'' \emph{Class. Quant. Grav.}, vol.~24, pp. 4533--4538, 2007.

\bibitem{SevillanoMunoz:2022tfb}
S.~Sevillano Mu\~noz, E.~J. Copeland, P.~Millington, and M.~Spannowsky, ``{FeynMG: A FeynRules extension for scalar-tensor theories of gravity},'' \emph{Comput. Phys. Commun.}, vol. 296, p. 109035, 2024.

\bibitem{Burrage:2023rqx}
C.~Burrage and P.~Millington, ``{Higgs-induced screening mechanisms in scalar-tensor theories},'' \emph{Annals N. Y. Acad. Sci.}, vol.~00, pp. 1--9, 2023.

\bibitem{Wang:2012kj}
J.~Wang, L.~Hui, and J.~Khoury, ``{No-Go Theorems for Generalized Chameleon Field Theories},'' \emph{Phys. Rev. Lett.}, vol. 109, p. 241301, 2012.

\bibitem{Metz:2020vxd}
A.~Metz, B.~Pasquini, and S.~Rodini, ``{Revisiting the proton mass decomposition},'' \emph{Phys. Rev. D}, vol. 102, p. 114042, 2020.

\bibitem{Alarcon:2011zs}
J.~M. Alarcon, J.~Martin~Camalich, and J.~A. Oller, ``{The chiral representation of the $\pi N$ scattering amplitude and the pion-nucleon sigma term},'' \emph{Phys. Rev. D}, vol.~85, p. 051503, 2012.

\bibitem{Alarcon:2012nr}
J.~M. Alarcon, L.~S. Geng, J.~Martin~Camalich, and J.~A. Oller, ``{The strangeness content of the nucleon from effective field theory and phenomenology},'' \emph{Phys. Lett. B}, vol. 730, pp. 342--346, 2014.

\bibitem{Hoferichter:2015hva}
M.~Hoferichter, J.~Ruiz~de Elvira, B.~Kubis, and U.-G. Mei\ss{}ner, ``{Roy\textendash{}Steiner-equation analysis of pion\textendash{}nucleon scattering},'' \emph{Phys. Rept.}, vol. 625, pp. 1--88, 2016.

\bibitem{PhysRevD.102.054517}
C.~Alexandrou, S.~Bacchio, M.~Constantinou, J.~Finkenrath, K.~Hadjiyiannakou, K.~Jansen, G.~Koutsou, and A.~V. Aviles-Casco, ``Nucleon axial, tensor, and scalar charges and $\ensuremath{\sigma}$-terms in lattice qcd,'' \emph{Phys. Rev. D}, vol. 102, p. 054517, Sep 2020.

\bibitem{Yang:2018nqn}
Y.-B. Yang, J.~Liang, Y.-J. Bi, Y.~Chen, T.~Draper, K.-F. Liu, and Z.~Liu, ``{Proton Mass Decomposition from the QCD Energy Momentum Tensor},'' \emph{Phys. Rev. Lett.}, vol. 121, no.~21, p. 212001, 2018.

\bibitem{Burrage_2015}
C.~Burrage, E.~J. Copeland, and E.~Hinds, ``Probing dark energy with atom interferometry,'' \emph{Journal of Cosmology and Astroparticle Physics}, vol. 2015, no.~03, pp. 042--042, mar 2015.

\bibitem{Elder_2016}
B.~Elder, J.~Khoury, P.~Haslinger, M.~Jaffe, H.~Müller, and P.~Hamilton, ``Chameleon dark energy and atom interferometry,'' \emph{Physical Review D}, vol.~94, no.~4, aug 2016.

\bibitem{AION}
L.~Badurina \emph{et~al.}, ``{AION: An Atom Interferometer Observatory and Network},'' \emph{JCAP}, vol.~05, p. 011, 2020.

\bibitem{Damour:2010rm}
T.~Damour and J.~F. Donoghue, ``{Phenomenology of the Equivalence Principle with Light Scalars},'' \emph{Class. Quant. Grav.}, vol.~27, p. 202001, 2010.

\bibitem{Damour:2010rp}
------, ``{Equivalence Principle Violations and Couplings of a Light Dilaton},'' \emph{Phys. Rev. D}, vol.~82, p. 084033, 2010.

\bibitem{Copeland:2006wr}
E.~J. Copeland, M.~Sami, and S.~Tsujikawa, ``{Dynamics of dark energy},'' \emph{Int. J. Mod. Phys. D}, vol.~15, pp. 1753--1936, 2006.

\bibitem{Copeland:1997et}
E.~J. Copeland, A.~R. Liddle, and D.~Wands, ``{Exponential potentials and cosmological scaling solutions},'' \emph{Phys. Rev. D}, vol.~57, pp. 4686--4690, 1998.

\bibitem{Weinberg:1988cp}
S.~Weinberg, ``{The Cosmological Constant Problem},'' \emph{Rev. Mod. Phys.}, vol.~61, pp. 1--23, 1989.

\bibitem{Martin:2012bt}
J.~Martin, ``{Everything You Always Wanted To Know About The Cosmological Constant Problem (But Were Afraid To Ask)},'' \emph{Comptes Rendus Physique}, vol.~13, pp. 566--665, 2012.

\bibitem{2021MNRAS.506.5721A}
J.~{An}, A.~P. {Naik}, N.~W. {Evans}, and C.~{Burrage}, ``{Charting galactic accelerations: when and how to extract a unique potential from the distribution function},'' vol. 506, no.~4, pp. 5721--5730, Oct. 2021.

\bibitem{Brookfield:2005bz}
A.~W. Brookfield, C.~van~de Bruck, D.~F. Mota, and D.~Tocchini-Valentini, ``{Cosmology of mass-varying neutrinos driven by quintessence: theory and observations},'' \emph{Phys. Rev. D}, vol.~73, p. 083515, 2006, [Erratum: Phys.Rev.D 76, 049901 (2007)].

\bibitem{Sakstein:2019fmf}
J.~Sakstein and M.~Trodden, ``{Early Dark Energy from Massive Neutrinos as a Natural Resolution of the Hubble Tension},'' \emph{Phys. Rev. Lett.}, vol. 124, no.~16, p. 161301, 2020.

\bibitem{Casas:2024xqy}
G.~F. Casas, M.~Montero, and I.~Ruiz, ``{Cosmological Chameleons, String Theory and the Swampland},'' 6 2024.

\bibitem{2017MNRAS.465...76M}
P.~J. {McMillan}, ``{The mass distribution and gravitational potential of the Milky Way},'' vol. 465, no.~1, pp. 76--94, Feb. 2017.

\bibitem{Ruan:2021wup}
C.-Z. Ruan, C.~Hern\'andez-Aguayo, B.~Li, C.~Arnold, C.~M. Baugh, A.~Klypin, and F.~Prada, ``{Fast full N-body simulations of generic modified gravity: conformal coupling models},'' \emph{JCAP}, vol.~05, no.~05, p. 018, 2022.

\bibitem{Hernandez-Aguayo:2021kuh}
C.~Hern\'andez-Aguayo, C.-Z. Ruan, B.~Li, C.~Arnold, C.~M. Baugh, A.~Klypin, and F.~Prada, ``{Fast full N-body simulations of generic modified gravity: derivative coupling models},'' \emph{JCAP}, vol.~01, no.~01, p. 048, 2022.

\bibitem{2016PhRvL.117t1101M}
S.~S. {McGaugh}, F.~{Lelli}, and J.~M. {Schombert}, ``{Radial Acceleration Relation in Rotationally Supported Galaxies},'' vol. 117, no.~20, p. 201101, Nov. 2016.

\bibitem{Burrage:2016yjm}
C.~Burrage, E.~J. Copeland, and P.~Millington, ``{Radial acceleration relation from symmetron fifth forces},'' \emph{Phys. Rev. D}, vol.~95, no.~6, p. 064050, 2017, [Erratum: Phys.Rev.D 95, 129902 (2017)].

\bibitem{Grana:2021zvf}
M.~Gra\~na and A.~Herr\'aez, ``{The Swampland Conjectures: A Bridge from Quantum Gravity to Particle Physics},'' \emph{Universe}, vol.~7, no.~8, p. 273, 2021.

\bibitem{Vafa:2005ui}
H.~Ooguri, E.~Palti, G.~Shiu, and C.~Vafa, ``{Distance and de Sitter Conjectures on the Swampland},'' \emph{Phys. Lett. B}, vol. 788, pp. 180--184, 2019.

\end{thebibliography}
\end{document}